\documentstyle[epsf]{article}
\renewcommand{\theequation}{\mbox{\arabic{section}.\arabic{equation}}}

\begin{document}

\def\p{\phi}
\def\P{\Phi}
\def\a{\alpha}
\def\e{\varepsilon}
\def\be{\begin{equation}}
\def\ee{\end{equation}}
\def\l{\label}
\def\0{\setcounter{equation}{0}}
\def\b{\beta}
\def\S{\Sigma}
\def\C{\cite}
\def\r{\ref}
\def\ba{\begin{eqnarray}}
\def\ea{\end{eqnarray}}
\def\n{\nonumber}
\def\R{\rho}
\def\X{\Xi}
\def\x{\xi}
\def\la{\lambda}
\def\d{\delta}
\def\s{\sigma}
\def\f{\frac}
\def\D{\Delta}
\def\pa{\partial}
\def\Th{\Theta}
\def\o{\omega}
\def\O{\Omega}
\def\th{\theta}
\def\ga{\gamma}
\def\Ga{\Gamma}
\def\h{\hat}
\def\rar{\rightarrow}
\def\vp{\varphi}
\def\le{\left}
\def\ri{\right}
\def\foot{\footnote}

\begin{center}
{\Large\bf Fields topology and observables}
\vskip 0.2cm

{\large\it J.Manjavidze\foot{Permanent address: Inst. of Physics,
Tnilisi, Georgia. E-mail: joseph@nu.jinr.ru}, A.Sissakian}\\ JINR,
Dubna, Russia

\end{center}

\begin{abstract}

The paper contains successive description of the strong-coupling
perturbation theory. Formal realization of the idea
is based on observation that the path-integrals measure for
absorption part of amplitudes $\R$ is Diracian ($\d$-like).
New form of the perturbation theory achieved mapping the quantum
dynamics in the space $W_G$ of (action, angle) type collective
variables.  It is shown that the transformed perturbation theory
contributions are accumulated on the boundary $\pa W_G$, i.e.  are
sensible to the $topology$ of factor space $W_G$ and,therefore, to
the theory symmetry group $G$.
The abilities of our perturbation theory are illustrated
by examples from quantum mechanics and field theory. Considering the
Coulomb potential $1/|x|$ the total reduction of quantum degrees of
freedom is demonstrated mapping the dynamics in the (angle, angular
momentum, Runge-Lentz vector) space. To solve the
(1+1)-dimensional sin-Gordon model the theory is considered in the
space (coordinates, momenta) of solitons. It is shown the total
reduction of quantum degrees of freedom and, in result, there is not
multiple production of particles.
This result we interpret as the $S$-matrix form of confinement. The
scalar $O(4,2)$-invariant field theoretical model is quantized in the
$W_O=O(4,2)/O(4) {\times} O(2)$ factor manifold.  It is shown that
the corresponding $\R$ is nontrivial because of the scale invariance
breaking.

\end{abstract}

\section{Introduction.}\setcounter{equation}{0}

The mostly intriguing phenomena in modern particles physics is
unobservability of color charge in a free state. The underling vector
fields theory posses the conformal $O(4,2)$ symmetry supported by the
non-Abelian gauge symmetry.  Absence of quantitative theory of the
confinement leads to enormous number of speculations both in
particles physics and in cosmology, e.g. \C{wilch}. To all appearance
the theory of this phenomena may have influence on the ordinary
statistics also.

Solution of the problem of confinement rest on absence of workable
perturbation theory. The mostly powerful WKB expansion (or the
stationary phase method in the functional integral terms) is
noneffective if the fields topology is nontrivial \C{wkb} since the
dynamics of perturbations of such fields is rather complicated.
Beside that, to describe the renormalized vacuum of highly
symmetrical modern systems, i.e.  non-Abelian gauge fields, the
infinite number of order parameters may be essential to find elements
of $S$-matrix\footnote {There is many quantitative attempts to
estimate the set of needed polynomial over fields order parameters to
describe the spectra of classical hadrons.  The to-day situation in
this field is described in \C{shif}. There is also an attempt to
introduce highly nonlinear over fields order \C{pol, svet} (or
disorder \C{t'h}) parameter.  See also \C{ellis}, where the
phenomenological aspects of last one are discussed.}.\\

$\bullet$ The main goal of this paper is to present the perturbation
theory to handle discussed problem. Our perturbation theory is
nothing new if the topology of interacting fields is trivial, but is
extremely effective for nontrivial topology case. Actually we offer
the successive approach to the strong-coupling perturbation theory.
Few examples from quantum mechanics (the Coulomb problem is solved
exactly) and ((1+1)-dimensional sin-Gordon, scalar $O(4,2)$-
invariant) field theories will be examined to illustrate abilities of
the approach.

The aim of this review paper is to show

(i)  {\it The origin of desired perturbation theory.}

Formally the approach based on the remark \C{yad} that the
path-integral representation just of observables is defined on the
Dirac measure
$$\prod_x du(x)\d (\pa_\mu^2 u+v'(u)-j),$$
where $j(x)$ is the random, defined on Gauss measure, source of
quantum perturbations. We would show the mechanism of unitary, i.e.
the total probability conserving, canonical mapping, see Sec.5, of
the functional measure on the factor space $W_G=G/\bar{G}$, where $G$
is the theory symmetry group and $\bar{G}$ is the symmetry of
classical fields $u_c=u_c(x;\x,\eta)$.  In other words, we would like
to describe the quantum dynamics in the space of field parameters
$(\x,\eta)(t)$.  For instance, see Sec.9, $W_G$ is the space of
coordinates $\x$ and momenta $\eta$ of solitons.  The coordinate
transformations $x_\mu\rar x_\mu'=X_\mu(x)$ will be described also,
see Sec.6. In this case the field parameters would define the inner
geometry of space with definite metrics, see also Secs.6,9.

(ii) {\it The structure of perturbation theory in the $G/\bar{G}$
space and as it can be applied.}

We want to warn that a reader may meet new for quantum theories
phenomena in the developed strong-coupling perturbation
theory\footnote {The probably noncompleat description of today
experience in this field will be given in Sec.2.}.  By the same
reason some conclusion would be seem unusual. It will be shown, see
Secs.5,7,8,9, that the quantum corrections of the transformed theory
are accumulated on the boundaries (bifurcation manifolds \C{<<<})
$\pa W_G$ of the factor space, i.e.  are defined mainly by its
$topology$.  This and $\d$-likeness of measure for $observables$
justify the title of the paper.

The main quantitative consequence would be the observation that the
quantum corrections may disappeared (totally or partly) on $\pa W_G$.
I.e., the problem of quantum corrections we reduce up to definition
of intersection $\{\pa u_c\} \bigcap \pa W_G$ of the defining
interactions set $\{\pa u_c(\x,\eta)\}$ with the boundary $\pa W_G$.
This is the new phenomena of reduction of the quantum corrections to
the quasiclassical approximation;

(iii) {\it The range of validity of described approach.}

We would discuss the general form of generating functional of
observables, see Secs.2,3, to eliminate doubts in generality of the
approach. This question seems important since we would formulate
formalism in terms of $observables$. The Dirac measure of
functional-integral representation of observables unambiguously
defines the complete set of states of the interacting fields. This
allows to construct the perturbation theory in arbitrary useful
terms. But, performing transformation, it is impossible come back to
the theory in terms of amplitudes since the transformation mixes
various degrees of freedom and there is not factorization of $\R$ on
the product of two amplitudes. We are obliged in result to deal with
the observables, or theirs generating functional $\R$. This was the
reason why the question of generality of our approach was considered
in Secs.3,4.\\

$\bullet$ The paper is organized as follows.

-- Sec.2. The overview of main ideas and results will be described
qualitatively.  We hope that this additional introductory section
will help to read this large paper.

-- Sec.3. The generating functional of observables $\R$ will be
introduced.  It will be shown that $\R$ can be used both for generation
of observables in particles physics and for description of classical
statistical media.

-- Sec.4. The differential measure for $\R$ is derived and
it is shown that the approach is applicable iff the theory is infrared
stable. The Wigner functions formalism would by applied for this
purpose.

-- Sec.5. The main properties of new perturbation theory in the
invariant subspace (factor manifold) would be shown considering
simplest quantum-mechanical example.

-- Sec.6. The main purpose of this section is to show that the
transformed theory can not be deduced by $naive$ transformation of
integrands of initial path integral for amplitudes. This question is
important searching the functional measure on the group manifold.

-- Sec.7. We consider the Coulomb problem to show explicitly the role
of additional reduction for quantum systems. This phenomena will be
used describing the scalar $O(4,2)$-invariant field theory.

-- Sec.8. The sin-Gordon model is simplest integrable
field-theoretical models. By this reason this model will be
considered to calculate $\R$ using the mapping on the invariant
subspace. The explicit calculation leads to trivial $\R=1$, as the
consequence of sin-Gordon solitons stability (or, in other words, of
the $\tilde{sl}(2,C)$ Kac-Moody algebra unbrokeness)

-- Sec.9. The $O(4,2)$ model will be considered to demonstrate
explicitly the mapping of the 4-dimensional nonintegrable field
theory on the factor manifold. It will be shown that $O(4,2)$
symmetry is broken up to $O(4){\times}O(2)$ symmetry. This phenomena
of the scale invariance breaking gives the nontrivial $\R$, in
contrast to the sin-Gordon model.

-- Sec.10. We observe the main details of the approach.

For sake of convenience we extract through the text of this paper the
main steps as the statements\footnote{Some of them are evident and
the proof of others is given with a `physical' level of strictness.}.
The cumbersome calculations are shifted in the Appendices.

\section{Introductory review.}\setcounter{equation}{0}

It seems useful, noting large volume of the paper, to give
previously the qualitative review of its content. The quantitative
realizations, considering simplest characteristic examples, are given
in subsequent sections.\\

$\bullet$  As was underlined above, the particularity of the approach
consist in direct calculation of the observables leaving behind
calculation of amplitudes. By this reason the useful
functional-integral representation for observables will be discussed,
specially concentrating attention on the universality of this
description.

So, we will find the order parameters substitute natural for
$S$-matrix theory. The parameter $\Ga(q;u)$ directly connected with
$external$ particles energy, momentum, spin, polarization, charge,
etc., and sensitive to symmetry properties of the interacting
fields system will be introduced\footnote{Following trivial analogy
with ferromagnetic may be useful. So, the external magnetic field
$\cal{H}$$\sim \bar{\mu}$, if $\bar{\mu}$ is the magnetics order
parameter, and the phase transition means that $\bar{\mu}\neq 0$.
Offered $\Ga (q,u)$ has the same meaning as $\cal{H}$.}. For sake of
simplicity we will consider in this paper $u(x)$ as the real scalar
field. The generalization would be evident.

In used $S$-matrix approach the expectation value of $\Ga(q;u)$ will
define the asymptotic states density. So, the $m$- into $n$-particles
transition (nonnormalized) $probability$ $\R_{nm}$ would have a
symmetrical form in terms of $\Ga(q;u)$:
\ba
\R_{nm}(q_1,...,q_n;q'_1,...,q'_m)=
\n\\
<\prod^{n}_{k=1}|\Ga(q_k;u)|^2 \prod^{m}_{k=1}|\Ga(q'_k;u)|^2>_u.
\l{1.1}\ea
Here $q'(q)$ are the in(out)-going particles momenta and $<>_u$ means
averaging over field $u(x)$.

By definition, $\R_{2n}(q'_1, q'_2;q_1,..., q_n)$ can be used as the
generator of events modeling the accelerator experiments \C{siss,
elpat}. For instance,
$$
\f{1}{J(s)}\sum_n\int_{s}
\R_{2n}(q_1,...,q_n;q'_1,q'_2)= \s_{tot}(s),
$$
where $\s_{tot}$ is the total cross section ($J$ is the ordinary
normalization factor). In this expression integrations are
performed with constraint $s=(q'_1+q'_2)^2$.

Beside that, the ordinary big partition function of statistical
system can be expressed through
\be
\sum_{n,m}\int_{(\b_i,z_i;\b_f,z_f)}
\R_{nm}(q_1,...,q_n;q'_1,...,q'_m),
\l{densm}\ee
where $\R_{nm}$ is defined by (\r{1.1}). The summation and
integration are performed with constraints that the mean energy of
particles in initial(final) state is $1/\b_i(1/\b_f)$. So, $1/\b$ has
the meaning of temperature and $z_i(z_f)$ is the activity for
initial(final) state. We investigate in what conditions such
description of statistics is rightful. It will be seen that the
Bogolyubov's relaxation of correlations near equilibrium is the only
condition \C{elpat}.

It will be shown that this definition of statistics has the right
classical limit iff the theory is infrared stable, i.e. iff the
4-dimensional scale of quantum fluctuations $L_q$ is much smaller
then the scale of statistical ones $L_s$. In this case one can
deduce from $\R_{nm}$ the classical phase space $(R,q)$ distribution.
The Wigner-functions $W(R,q)$ formalism \C{wig} will be
applied, see also \C{carr} to show the origin of $L_q<<L_s$
condition.  The details one can find in Sec.3\footnote{There is also
the `marginal distribution functions' formalism \C{man'ko} as the
attempt of universal description both quantum and statistical
systems. It will be seen that the Wigner-functions approach is
natural for $S$-matrix \C{carr} and gives the evident physical
interpretation of transition from quantum to classical physics.}.\\

$\bullet$. We will find that $\Ga(q;u)$ is the function of external
particles momentum $q$ and is the linear functional of $u(x)$:
\ba
\Ga(q;u)=-\int dx e^{iqx} \f{\d S_0 (u)}{\d u(x)}=
\n \\
\int dx
e^{iqx}(\pa^2 +m^2)u(x) ,~~q^2=m^2,
\l{1.2}\ea
for the mass $m$ field. Here $S_0(u)$ is the free part of the
action. This parameter presents the momentum distribution of the
interacting field $u(x)$ on the remote hypersurface $\s_\infty$ if
$u(x)$ is the regular function. $Note$, the operator $(\pa^2 +m^2)$
cancels the mass-shell states of $u(x)$.

$Note$, the definition (\r{1.1}) is rightful, i.e. $q$ is the external
particles momentum and the condition $q^2=m^2$ is hold, iff the
theory is infrared stable. It should be underlined, see Sec.4, that
this condition is `seen' if the $(R,q)$ distribution in the classical
limit is considered.

The construction (\r{1.2}) means, because of the Klein-Gordon operator
and the external states should be mass-shell by definition \C{lan},
the solution $\R_{nm}=0$ is possible for particular topology
(compactness and analytic properties) of $quantum$ field $u(x)$. So,
$\Ga(q;u)$ carry following remarkable properties:

--it directly defines the observables,

--is defined by the topology of $u(x)$,

--is the linear functional of the actions symmetry group element
$u(x)$.

$Note$, the space-time topology of $u(x,t)$ becomes important
calculating integral (\r{1.2}) by parts. This procedure is available
iff $u(x,t)$ is the regular function. But the $quantum$ fields are
always singular since Green functions are the distributions (see
Sec.8). Therefore, the solution $\Ga(q;u)=0$ is valid iff the
quasiclassical approximation is exact. Just this situation is
realized in the soliton sector of sin-Gordon model.\\

$\bullet$ It is evident that if $\R_{nm}=0$, $n,m\neq 0$, the field
$u(x)$ is confined since is not measurable as the external field.
This is the $S$-matrix interpretation of the `confinement' phenomena.
Just this formulation of the confinement we would like to develop in
future.  $Note$, $\R_{nm} \neq 0$ testify not only the broken
symmetry but this expectation value is not zero in absence of any
symmetry also.

Despite this ambiguity $\Ga(q;u)$ carry the definite properties of
the order parameter since the opposite solution $\R_{nm}=0$ can be
the dynamical display of an $unbroken$ symmetry only\footnote{The
$S$-matrix would be introduced phenomenologically, postulating the
ordinary Lehmann-Symanzik-Zimmermann (LSZ) reduction formulae, see
eq.(\r{2.1'}). So, the formal constraints, e.g. the Haag theorem,
would not be taken into account.}, i.e. of the nontrivial topology of
interacting fields, as the consequence of unbroken symmetry.

Note `sparingness' of the $S$-matrix description. The external
states of the $S$-matrix approach should be expandable on the Fock
basis and belong to the mass shell. So, we can ask only -- `is
this particle (or multiparticle state), with given properties,
observable in the free state'.  We hope to construct the perturbation
theory which is able to give the answer on this question
unambiguously for arbitrary Lagrangian in arbitrary space-time
dimension.\\

$\bullet$ Our hope based on the particularity of offered below
quantization method. It consist in introduction of the unitarity
condition in the path-integral formalism \C{yad}. It will be seen
that the dynamical display of unitarity condition is the local
equilibrium of all forces in the quantum system. In other words, the
total probability conservation principle is the quantum analogy of
the classical D'Alembert's variational principle \C{yaph}.  This
$dynamical$ scheme seems interesting if one remained above
mentioned problem with description of modern systems vacuum, it
allows to perform calculations without going into details of the
renormalized vacuum structure.

So, it will be shown (see also \C{yad}) that the unitarity
condition leads to $\d$-like\footnote{In the mathematical literature
such measure is known as the `Dirac measure'.} functional measure
$DM(u)$ for $\R_{nm}$:
\be
DM(u)=\prod_x du(x) \d \left( \f{\d S(u)}{\d u(x)}-j(x)\right) ,
\l{a'}\ee
where $j(x)$ is the provoking quantum excitations force and $S(u)$ is
the classical action. It is evident that this measure close logically
the formalism since it defines a complete set of contributions for
given classical Lagrangian in the sector of real-time fields.

One should assume that $j(x)$ switched on adiabatically (in this case
we expand contributions in the vicinity of $j=0$) for effective use
of this definition of measure.  Otherwise we should know $j(x)$
exactly, including it into Lagrangian as the external field. The
measure would remain $\d$-like in last case also. $Note$, $j(x)$ can
be introduced adiabatically even if the dynamical symmetry breaking
is expected \C{col}.

So, the measure (\r{a'}) allows to conclude that the {\it sum of all
(including trivial), strict, regular, real-time} solutions of the
classical equation \C{yaph}
\be
\f{\d S(u)}{\d u(x)}=0.
\l{equ}\ee
defines the complete set of contributions.

$Note$, eq.(\r{equ}) reflects the ordinary Hamilton variational
principle if the quantum perturbations switched on adiabatically.
We would show therefore that the WKB expansion is in agreement with
unitarity condition. It is the well known result. But the measure
(\r{a'}) contains following new information:

{\bf a.} Only $strict$ solutions of eq.(\r{equ}) should be taken into
account. This `rigidness' of the formalism means absence of
pseudo-solutions (similar to multi-instanton, or multi-kink)
contribution.

{\bf b.} $\R_{nm}$ is described by the $sum$ of all solutions of
eq.(\r{equ}), independently from theirs `nearness' in the functional
space;

{\bf c.} $\R_{nm}$ did not contain the interference terms from various
topologically nonequivalent contributions. This displays the
orthogonality of corresponding Hilbert spaces;

{\bf d.} The measure (\r{a'}) includes $j(x)$ as the external source.
Its fluctuation disturb the solutions of eq.(\r{equ}) and {\it vice
versa} since the measure (\r{a'}) is strict;

{\bf e.} In the frame of above adiabaticity condition the field
disturbed by $j(x)$ belongs to the same manifold (topology class) as
the classical field defined by (\r{equ}) \C{yaph}\footnote{There are
in modern physics the remarkable attempts to construct a geometrical
approach to quantum mechanics \C{kibble, anandan} and field theory
\C{konopl}. Our approach, based on the dynamical equilibrium, see
(\r{a'}), between classical and quantum forces, contains evident
geometrical interpretation and it will be widely used, see below.
$Note$, our interpretation have deal with the classical phase-space
flows. By this reason, in contrast with above mentioned approaches,
the finite dimensional manifolds only, as in classical mechanics,
would arise.}.

One must underline that the measure (\r{a'}) is derived for
$real-time$ processes only, i.e.  is not valid for tunneling ones. By
this reason above conclusions should be taken carefully. The
corresponding selection rule will be given below in Sec.5.

$Note$, the $\d$-likeness of measure essential to find eq.(\r{1.2}).
This will be shown in Sec.4.

$Note$ also that some information is lost calculating probabilities
only. We will show as the approach can be generalized to calculate
the imaginary part of the elastic scattering amplitudes.  Then, using
dispersion relation one can restore the total amplitudes.\\

$\bullet$ We start discussion of the approach, risking to loose
generality, by simplest quantum-mechanical examples of particle
motion in the potential hole $v(x)$ with one non-degenerate minimum
at $x=0$. We will calculate the probability $\R=\R (E)$ to find the
bound state with energy $E$.

It will be seen that this experience is useful for quantization of
nonlinear waves also. Indeed, introducing the convenient variables
(collective coordinates) one can reduce the quantum soliton-like
excitations problem to quantum-mechanical one. This idea was
considered previously by many authors, e.g. \C{gold, kor}.

Quantitatively the problem looks as follows. It is not difficult to
describe the one-particles dynamics in the quasiclassical
approximation since corresponding equation for trajectory $x_c$
always can be solved. But, beyond this approximation, to use the
ordinary WKB expansion of path integrals, one should solve the
equation for Green function $G(t,t')$:  \be (\pa^2 + v''(x_c))_t
G(t,t')=\d (t-t').  \l{2}\ee Just eq.(\r{2}) offers a difficulty: it
is impossible to find an exact solution of this equation since $x_c
=x_c(t)$ is the non-trivial function and, therefore, $G(t,t')$ is not
translationally invariant describing propagation of a `particle' in
the $time- dependent$ potential $v''(x_c)$\footnote{One can find
$G(t,t')$ perturbatively, expanding $v''(x_c)$ over $x_c$. However,
it is too hard to handle the resulting double-parametric perturbation
theory.}.  We avoid this difficulty introducing the convenient
dynamical variables.

The main formal difficulty, see e.g. \C{mar}, of this program
consisting in transformation of the path-integral measure was solved
in \C{yad}\foot{Number of problems of
quantum mechanics was solved using also the `time sliced' method
\C{groshe}. This approach presents the path integral as the finite
product of well defined ordinary integrals and, therefore, allows
perform arbitrary space and space-time transformations. But
transformed `effective' Lagrangian gains additional term
$\sim\hbar^2$.  Last one crucially depends from the way as the
`slicing' was performed. This phenomena considerably complicate
calculations and the general solution of this problem is unknown
for us. It is evident that this method is effective if the
quantum corrections $\sim\hbar$ play no role.  Such models are well
known.  For instance, the Coulomb model in quantum mechanics, the
sine-Gordon model in field theory, where the bound-state energies are
exactly quasiclassical.}:  the phase space path-integrals
differential measure $DM(x,p)$ for $\R$ has the form (see (\r{a'})):
\be
DM(x,p)\sim\prod_t dx(t)dp(t)\d (\dot{x} +
\frac{\pa H_j}{\pa p}) \d (\dot{p} - \frac{\pa H_j}{\pa x}),
\l{3}\ee
where the Hamiltonian $$ H_j =\frac{1}{2}p^2 + v(x) -jx $$ includes
the energy of quantum fluctuations $jx$, with the provoking quantum
excitations force $j=j(t)$. Then the dynamical equilibrium between
ordinary mechanical forces (kinetic $\dot{p}(t)$ plus potential
$v'(x)$) and quantum force ($j(t)$) determined by $\d$-like measure
(\r{3}) allows to perform an arbitrary transformation of quantum
measure caused by transformation of classical forces, i.e. of $x$ and
$p$ (see {\bf d.}).

We will use this property introducing the `motion' on the cotangent
bundle $T^*G=(\th ,h)$, where $h$ is the bundles parameter and $\th$
is the coordinate on it. For definiteness, let $h$ be the conserved
on the $classical$ trajectory energy and $\th$ is the conjugate to
$h$ `time'. (The transformation to action-angle variables will be
described also.) The mapping $(x,p) \rightarrow (\th ,h)$ is
canonical and the corresponding equations of motion on the cotangent
bundle should have the form:
\ba
\dot{h}=-\frac{\pa h_j}{\pa\th}=+j\frac{\pa x_c (\th ,h)}{\pa \th},
\n \\
\dot{\th}=+\frac{\pa h_j}{\pa h}=1-j\frac{\pa x_c (\th ,h)}{\pa h},
\l{4}\ea
where
$$
h_j (\th ,h) =h -jx_c (\th ,h)
$$
is the transformed Hamiltonian and $x_c (\th ,h)$ is the classical
trajectory in the $(\th ,h)$ terms. The Green function of the
eq.(\r{4}) $g(t,t')$ is translationally invariant since classically
(at $j=0$) the cotangent bundle is the time-independent manifold.

Above example shows that the mapping is constructive iff the bundle
parameters are generators of (sub)group violated by $x_c$.
Corresponding phase space is the invariant subspace \C{smale} (see
also footnote 8).

For some problems the mapping on the curved space with nontrivial
metric tensor $g_{\mu\nu}(x)$ is useful, see, e.g., \C{dow}. We will
find the quantum mechanical Dirac measure directly in the curved
space with nontrivial $g_{\mu\nu}(x)$.

We examine the factorizability of transformed
measure for $\R$ considering transformation of the Dirac measure to
cylindrical coordinates. We will consider the general coordinate
transformation $x_\mu\rar x_\mu'=X_\mu(y)$ starting from flat space,
where the metric tensor $g_{\mu\nu}$ is trivial. The Dirac
measure for theory with Lagrangian
\be
L=\f{1}{2}g_{\mu\nu}(y)\dot{y}^\mu\dot{y}^\nu-v(y),
\l{l}\ee
where $g_{\mu\nu}(y)$ is the nontrivial function is calculated also
in Sec.6.\\

$\bullet$ It will be shown that the procedure of averaging $<>_u$ in
(\r{1.1}) carried out by the integro-differential operator
\be
\hat{\cal O}=e^{-i\h{K}(j,e)}\int DM(x)e^{-iU(x,e)}.
\l{r1}\ee
It should be underlined that this representation is strict and is
valid for arbitrary Lagrange theory of arbitrary dimensions. It can
be considered also as the `unitary definition' of the functional
integral.

The operator $\hat{\cal O}$ contains three element. The Dirac
measure $DM$, the functional $U(x,e)$ and the operator $\h{K}(j,e)$.
The functional $U(x,e)=O(e^3)$ describes interaction and is simply
calculable from $S(x)$. The operator $\h{K}(j,e)$ contains the
product $\h{j}\cdot\h{e}$ of the functional derivatives over $j$ and
$e$, i.e.  is Gaussian.  The expansion over $\h K$ gives perturbation
series. At the very end of calculations one should take the auxiliary
variables $j,~e$ equal to zero. $Note$, the structure of operator
$\hat{\cal O}$ is form-invariant concerning transformation of
variables ($x$ in the configuration space, $(x,p)$ in the phase
space).

The perturbation theory with measure (\r{3}) on the cotangent bundles
has unusual properties \C{untr, ang} and few proposition concerning
perturbation theory in the invariant subspaces will be offered.

The main of them is possibility to split the `Lagrange' source
$j(t)\rar j_{W_G}$, see (\r{4}), onto set of sources of each
independent degree of freedom of the invariant subspace. For
instance, the splitting \be j(t)\rar (j_\th (t), j_h (t))
\l{split}\ee will be demonstrated. By this way the actually
Hamilton's  description is achieved in the invariant subspace.
$Note$, the transformation (\r{split}) induce \be e(t)\rar (e_\th
(t), e_h (t)) \l{split'}\ee leaving a structure of $U(x,e)$
unchanged. This remarkable property allows to describe interactions
between various degrees of freedom in the invariant subspace since in
result \be \h{j}\cdot\h{e}\rar\h{j}_h\h{e}_h+\h{j}_\th\h{e}_\th.
\l{split''}\ee

We will consider also the transformation to the cylindrical
coordinates $(r,\vp)$. In this case the Cartesian sources
$(j_{x_1},j_{x_2})\rar(j_r,j_\vp)$ (and
$(e_{x_1},e_{x_2})\rar(e_r,e_\vp)$).

$Note$, the probability $\R\sim <in|out><in|out>^*$ describes the
`closed-path' motion by definition. This means that the corresponding
classical action $S_\wp=\oint_\wp pdx-Hdt$, where $\wp$ is composed
from two $(in\rar out)$ and $out\rar in)$ independent pathes, is the
Poincare invariant, i.e.  $S_\wp=S_{\wp'},$ if the phase-space flows
through the closed path $\wp$ and $\wp'$ one coincide.  We will see
that new perturbation theory describes such variations of $\wp$ that
it enclose different phase-space tubes since the quantum
perturbations are unable to change the phase-space `liquids' density.

So, resulting perturbation theory describes fluctuations of
phase-space flow, induced by $(j_\th (t), j_h (t))$, through the
elementary cells $\d x_c\wedge \d q_c$, where $\d q_c$ is the tangent
to $x_c$ vector and $\d x_c\equiv e$ is the virtual variation from
$x_c$. For instance,
\be
\d x_c\wedge \d q_c=e_\th\f{\pa x_c}{\pa h}-e_h\f{\pa x_c}{\pa\th}
\l{flow}\ee
in terms of local coordinates $(\th,h)$. Each term of the
perturbation series is proportional to derivatives of the
nondegenerate 2-form $d\o=\d x_c\wedge \d p_c$ over $\th$ and/or $h$.

Then the equality
\be
d\o_i=(\d x_c\wedge\d q_c)_i=0
\l{qflow}\ee
means that there is not $quantum$ flow, i.e. quantum corrections, in
the $i$-th direction of the invariant subspace. $Note$, eq.(\r{qflow})
is hold if one of elements in the product ($(\d x_c)_i$ or $(\d
q_c)_i$) is equal to zero if the two-form $d\o$ is not degenerate in
the ordinary sense. This will allow to consider not only the
canonical transformations of the path-integral measure.

Secondly, each order of perturbation theory can be reduced to the sum
of total derivative over global coordinate of the invariant subspace.
So, if $(\th_0, h_0)$ are the initial data for classical trajectory
in the $(\th, h)$ space then the quantum corrections to the
quasiclassical approximation
\be
\R^q= \int d\th_0 dh_0 \{\f{\pa \R_{h}}{\pa h_0} + \f{\pa
\R_{\th}}{\pa \th_0}\}= \int_{\pa W_G}d\R^q.
\l{topol}\ee
This property is important analyzing integrability of quantum systems
since brings forward the topology of contributions, i.e. the boundary
properties of classical trajectory $x_c (\th,h)$, or, in other words,
the topology of factor manifold.

$Note$, the measure of integration in (\r{topol}) over zero modes
degrees of freedom ($\th_0$ in this case) was defined as the result of
mapping on the invariant subspace, i.e. without introduction of the
Faddeev-Popov $anzats$\foot{This can help to avoid the
problems with Gribov's ambiguities quantizing the non-Abelian
Yang-Mills theory. Above suggestion means that the gauge phase will
be considered as a dynamical degree of freedom, we suppose to
quantize the Yang-Mills theory in the $W_G{\times}\O_g$ space,
where $\O_g$ is the gage orbit.}.\\

$\bullet$ Above example shows that it is necessary to have the
universal method of mapping of quantum dynamics on the useful
manifolds. Let us introduce helpful notations. So, it is well known
that the mapping \be J:T \rar W, \l{redac}\ee where $T$ is the
$2N$-dimensional phase space and $W$ is a linear space, solves the
mechanical problem exactly iff \be J=\otimes^N_1 J_i, \l{iff}\ee
where $J_i$ are the first integrals in involution, e.g. \C{arn} (the
formalism of reduction (\r{redac}) in classical mechanics is
described also in \C{mars}).

In other words, if $J_i=J_i(x,p)$, $i=1,2,...,N$, are the first
integrals, $\dot{J}_i=0$, in involution, $\{J_i,J_j\}=0$, following
to Liouville-Arnold theorem, the $canonical$ transformation
$(x,p)\rar(J,Q)$ solves the mechanical problem. Then the $(x,p)_c$
flow is defined by the $2N$ system of coupled algebraic equations
$\eta=J(x,p),~\x=Q(x,p)$.

The mapping (\r{redac}) introduces integral $manifold$ $J_{\o}=
J^{-1} (\o)$ in such a way that the $classical$ phase space flaw with
given $\eta \in J_{\o}$ belongs to $J_{\o}$ $completely$ ($\eta
\equiv h_0$ in the above example).  Our methodological idea assumes
quantization of the $J_{\o}$ manifold instead of flow in $T$.

This becomes possible noting that the quantum trajectory should
belong to $J_{\o}$. Indeed, in the above example the trajectory
$x_c=x_c(\th,h)$ was defined by parameters $(\th,h)$ completely,
where $(\th,h)$ are defined by eqs.(\r{4}). $Note$, this important
property based on the $\d$-likeness of measure (see {\bf a.}) and
assumption that $j$ is switched on adiabatically.

The `direct' mapping (\r{redac}) used above, see \C{yaph}, assumes
that $J$ is known.  But this approach to the general problem of
mapping of quantum dynamics seems inconvenient having in mind the
nonlinear waves quantization, when the number of degrees of freedom
$N=\infty$, or if the transformation is not canonical (or if we do
not know is the transformation canonical).  We will consider by this
reason the `inverse' approach starting from assumption that the
classical flow is known.  Then, since (i) if the flow belongs to
$J_{\o}$ completely and (ii) if $J_{\o}$ presents the manifold
\C{yaph}, we would be able to introduce motion in $W$.

The manifold $J_{\o}$ is invariant relatively to some subgroup
$G_{\o}$ \C{smale} in accordance to topological class of classical
flaw\foot{By this reason $W$ is the invariant subspace.}. This
introduces the $J_{\o}$ classification and summation over all
(homotopy) classes should be performed.

$Note$, the classes are
separated by the boundary bifurcation lines in $W$ \C{smale}.  Then,
noting (\r{topol}), the quantum dynamics is defined on this
bifurcation lines.

If the quantum perturbations switched on adiabatically then the
homotopy group should stay unbroken. It is the ordinary statement for
quantum mechanics.

Therefore, we trying to describe in this paper the transition from
quantum dynamics in the trajectories space to the dynamics in the
space of parameters of classical trajectories. In the above example
just $(\th,h)$ were such parameters.

This step seems nontrivial since separates the problem of searching
of solutions of eq.(\r{equ}) and the problem of quantization in
corresponding vicinity. Then, noting that the quantum corrections are
accumulated on the boundary of factor manifold $\pa W$, we can
estimate, at least qualitatively, the contributions since definition
of $W$ is the algebraic problem, see Sec.9.

We would try to construct the mapping using the base of symplectic
geometry\foot{This idea is natural on the Dirac measure noting
that the canonical transformation is a one-parametric group of
diffeomorphysms.}.  This is possible having in mind discussion of
eq.(\r{qflow}).  Indeed, starting from assumption that $(\x,\eta)$
form the symplectic space (of arbitrary dimension) we would find the
measure projecting this space on $W$.  This step takes into account
the symmetry structure of the $W$ space (the unnecessary `empty'
degrees of freedom are canceled by the normalization condition).\\

$\bullet$ We will calculate the bound state energies in the Coulomb
potential to illustrate this idea. This popular problem was considered
by many authors, using various methods, see, e.g., \C{pop}. The
path-integral solution of this problem was offered in \C{dur} using
the time-sliced method.

We will restrict ourselves by the plane problem. Corresponding phase
space $T=(p,l,r,\vp)$ is 4-dimensional.

The classical flaw of this problem can be parametrized by the angular
momentum $l$, corresponding angle $\vp$ and by the normalized on
total Hamiltonian Runge-Lentz vectors length $n$. So, we will consider
the mapping ($p$ is the radial momentum in the cylindrical
coordinates):
\be
J_{l,n}:(p,l,r,\vp)\rar(l,n,\vp)
\l{15'}\ee
to construct the perturbation theory in the $W_C=(l,n,\vp)$ space.
I.e.  $W$ is not the symplectic space. Nevertheless we start from the
symplectic space adding to $n$ the auxiliary canonical variable $\x$.

Following to above definition, our transformed perturbation theory
describes the quantum flow through the oriented cell, see e.g.
(\r{flow}), in the invariant subspace. Then, following to (\r{15'}),
$d\o_n=(\d x_c\wedge\d p_c)_n=0$ (since $\pa r_c/\pa \x \equiv 0$)
and there is not quantum flow corresponding to $n$\footnote{So, $W_C$
would not have the symplectic structure for Coulomb potential. It
decomposed on the direct product $TG^* \times R^1$ of the
two-dimensional $(\vp ,l)$ phase space and ordinary one-dimensional
subspace $R^1$.  Last one correspond to $n$.  The quantum dynamics is
realized in the $(\vp ,l)$ phase space only.  $Note$, this conclusion
is in agreement with quantum uncertainty principle: if the dynamical
variable has not conjugate pare in the Poisson brackets sense, then
it stay being $c$-number in the quantization scheme. The integral
over $\x$ is canceled by normalization.}.  The Coulomb problem would
be considered to demonstrate just this partial reduction. We will
find that the bound-state energies are pure quasiclassical since the
quantum corrections are died out on the $\pa W_C=\pa (\vp,l)$
boundary.  \\

$\bullet$ We would consider the field theories with nontrivial
topology to verify the effectiveness of our formalism\foot{See also
footnote 3.}. For instance, we may consider creation of the free
particles in the theory with Lagrangian
\be
L=\f{1}{2}(\pa_{\mu}u)^2 + \f{m^2}{\la^2}[\cos(\la u)-1] .
\l{1.13}\ee
It is well known that this field model possess the soliton excitations
in the (1+1) dimension.

Formally nothing prevents to linearize partly our problem
considering the Lagrangian
\ba
L=\f{1}{2}[(\pa_{\mu}u)^2 - m^2 u^2] +
\f{m^2}{\la^2}[\cos(\la u)-1 + \f{\la^2}{2} u^2]
\n\\
\equiv S_0(u)-V(u)
\l{1.14}\ea
to describe particles creation (and absorption). The last term
describes interactions. Corresponding `order parameter' is
\be
\Ga (q;u)=\int dx
dt e^{iqx} (\pa^2 +m^2)u(x,t),~~~q^2 =m^2.
\l{or}\ee

We will calculate the expectation value:
$$
\R_{10}(q)\equiv\R_2 (q)=<|\Ga (q;u)|^2>_u.
$$
Just the procedure of averaging would be the object of our interest.
By definition, $\R_2$ is the probability to find one mass-shell
particle.  Certainly, $\R_2(q)=0$ on the sourceless vacuum but,
generally speaking, $\R_2(q)\neq 0$ in a field (soliton) with nonzero
energy density.

It will be shown by explicit calculations that
\be
\R_2(q) =0,~if~\vec{q} \neq 0,
\l{1}\ee
as the consequence of unbroken $\tilde{sl}(2,C)$ Kac-Moody algebra on
which the solitons of theory (\r{1.13}) are live, see e.g. \C{sol}
and references cited therein. The solution (\r{1}) seems important
since it can be interpreted as the explicit demonstration of field
$u(x,t)$ confinement\foot{This means also the solitons stability
under quantum perturbations. The same conclusion was established for
intagrable model \C{zamol} using the `$1/N$ expansion'.}. It will
be shown that (\r{1}) is the consequence of previously developed
proposition that the quasiclassical approximation is exact for
sin-Gordon model \C{dash}.

The same as for Coulomb problem reduction procedure will be applied.
But the reduction procedure of the sin-Gordon Hamiltonian system
with symmetry $G$ looks like canonical transformation \C{sol}, in
opposite to Coulomb problem, i.e. $W$ would have the symplectic
structure.  $Note$, considered mapping assumes the infinite number of
degrees of freedom reduction since it brings the quantum-field model,
defined in the infinite dimensional phase space $T$, up
to the quantum-mechanical one, defined in the finite dimensional
phase space $TG^*$ (see footnote 8).  We demonstrate this nontrivial
procedure explicitly.

So, we would show the way as the field-theoretical problem may be
reduced to the quantum-mechanical one. We would consider $\eta$ as
the `particles' (solitons) generalized momentum and would introduce
$\x$ as the conjugate to $\eta$ coordinate. The $N$-soliton
$2N$-dimensional phase space (cotangent manifold) $TG^*$ with local
coordinates $(\x,\eta)$ on it has natural symplectic structure, and
$DM(TG^*)=D^N M(\x,\eta)$ in practical calculations,
\be
D^N M(\x,\eta)=\prod_t \d(\dot{\x}-\f{\pa h_j}{\pa \eta})
\d(\dot{\eta}+\f{\pa h_j}{\pa \x}),
\l{equ'}\ee
where $h_j(\x,\eta)$ is the transformed total Hamiltonian and
$(\x,\eta)$ are the $N$ dimensional vectors. The summation over $N$
is assumed.

The quantum corrections to quasiclassical approximation of
transformed theory are simply calculable since the bundle parameters
$\eta$ are conserved in the classical limit.  This is the
particularity of solitons dynamics (solitons momenta are conserved
quantities). One can consider the developed formalism as the
path-integral version of nonlinear waves (solitons in our case)
quantum theory (the canonical quantization of sin-Gordon model in the
soliton sector was described also in \C{kor}.)

Noting that all solutions of model (\r{1.13}) are known \C{takh} we
will find $DM(TG^*)$ considering the mapping as an ordinary
transformation to useful variables.  This idea will be realized in
the following way. We can introduce following formal parametrization
of the field\footnote{The noncovariant notations $(x,t)$ are useful
since our perturbation theory is Lorentz-noncovariant.}
\be
u(x,t)=u_c(x;\x,\eta),
\l{param}\ee
where $(\x,\eta)(t)$ are the $N$-dimensional vectors. It is evident
that $u_c(x;\x,\eta)$ is the solution of incident equations iff it
obey the Poisson brackets ($p$ is the conjugate to $u$ momentum):
\be \{h_j,u\}=\f{\d H_j}{\d p},~~ \{h_j,p\}=-\f{\d H_j}{\d u},
\l{pois}\ee
at $u=u_c,~p =p_c$ and $(\x,\eta)$ obey the equations:
\be
\dot{\x}=\f{\pa h_j}{\pa \eta},~~
\dot{\eta}=-\f{\pa h_j}{\pa \x}.
\l{equa}\ee
$Note$, last equations are the arguments of $\d$-functions in the
transformed measure (\r{equ'}). \\

$\bullet$ The $O(4,2)$-invariant quantum field theory model with
Lagrangian
\be
L=\f{1}{2}(\pa_\mu u)^2-\f{1}{4}gu^4,
\l{11}\ee
where $u(x,t)$ is the real scalar field, beside its simplicity,
is important since the $ansatz$ for Yang-Mills potential \C{mil}
\be
B_\mu^a (x,t)= \eta_{\mu\nu}^a \pa^\nu \ln u (x,t)
\l{ansatz}\ee
reduces the Yang-Mills equations to the single one for the field $u$:
\be
\pa_\mu^2 u + gu^3 =0,~~g>0.
\l{12'}\ee
By this reason the theory (\r{11}) oftenly considered as a step toward
Yang-Mills theory, see also \C{act}.

We will calculate the expectation value
\be
\R_2(q)=<|\Ga (q,u)|^2>_u,
\l{13}\ee
where for given theory
\be
\Ga (q,u)=\int d^3x dt e^{iqx} \pa_\mu^2 u (x,t),~~q^2=0,
\l{14}\ee

Following to above experience, the $W$ space is the factor manifold
$(G/\bar{G})$ of the $G=O(4,2)$ group and $\bar{G}$ is the compact
\foot{Otherwise the classical solution would be singular.}
subgroup of $G$ group.  We will consider the highest $O(4){\times}
O(2)$ subgroup\foot{Another solutions of (\r{12'}) must be taken
into consideration also. So, strictly speaking,
considered in this paper contribution presents the particular
field-theoretical realization only.}.

So, let $\p(x,t)$ be the $O(4)\times O(2)$-invariant solution of
eq.(\r{12'}). Then the generators of translations $T_0$ and special
conformal transformations $K_0$ would define the coordinates of the
$W_O$ space\foot{The $O(4)\times O(2)$ subgroup of $O(4,2)$ group
is 7 parametric. So the factor manifold is 8 dimensional.}.

The classical phase flow in $W$ is simple:
\be
\dot{\x}_0=\o=const.,~~\dot{\eta}_0=0
\l{16'}\ee
The developed perturbation theory describes the random
fluctuations of $W$.

Following reduction procedure will be demonstrated. The functional
differential measure $DM(u, p)$ of the initial phase space
$V^\infty$ will be mapped on the space $W^8$ under the
phase flow $(\p, \pi)$, where $\pi=\dot{\p}$.  This reduction would
introduce the phase space with nondegenerate 2-form $\d \p \wedge
\d \pi$, where, as was explained above, $\d\p$ is the virtual
deviation of $u$ from the true trajectory $\p$ and $\d \pi$ is a
tangent to $\p$ vector of the transformed theory. This symplectic
structure allows perform further reduction (see discussion of
eq.(\r{qflow})): $W^8\rar W_O=T^*\bar{G}_1\times R^5$, where $R^5$ is
the 5-dimensional zero modes manifold and $T^*\bar{G}_1$ is the
4-dimensional phase space with one constraint\foot{This picture
reminds the Coulomb problem, see footnote 7.}.

All quantum dynamics is confined in $T^*\bar{G}_1$:
\be
\dot{\x}_i(t)=\o_i(\eta)+j_{\x_i},~~\dot{\eta}_i(t)=j_{\eta_i},~~i=1,2,
\l{17}\ee
with one constraint on the boundary conditions:
\be
\x_1(0)=\x_2(0).
\l{18}\ee
The sources $j_{\x_i},~j_{\eta_i}$ introduced for
description of the $T^*\bar{G}_1$ subspace quantum deformations.

Inserting into (\r{14}) the $O(4)\times O(2)$-invariant solution of
eq.(\r{12'}) \C{d'a, act} $\p (x,t)$ we find that
\be
\Ga (q,\p)=0,~~q_\mu \neq 0,
\l{15}\ee
since $\p (x,t) =0$ if $(x,t)$ belong to remote hypersurface
$\s_\infty$.  This example shows that the confinement condition
$\R_2(q)=0$ is hold in the quasiclassical approximation.

We investigate the influence of quantum corrections, i.e. the
question, can they alter the quasiclassical solution (\r{15}). It
will be shown that quantum corrections broke the scale invariance and
by this reason $\R_2(q)\neq 0$ for theory (\r{11}) on the $O(4)\times
O(2)$ -invariant solution.

$Note$, this example shows that the classical field $\p$ can broke
the symmetry, i.e. provoke the phase transition, in contrast to
kink-like excitations.

\section{Generating functional.}\0

For sake of completeness it seems useful to introduce from the very
beginning of this review paper the generating functional of observables
(cross sections, created particles spectra, etc.) $\R$. It seems also
interesting to investigate in a what conditions this quantity has the
right classical limit since in statistics $\R$ would have the meaning
of big partition function (see also \C{elpat}). In result we get to
the universal description of wide spectrum of physical phenomena.

This statement becomes evident remembering the microcanonical
approach in thermodynamics. We would like to  in this review
paper that (i) the distinction between thermodynamics and $S$-matrix
field theory consist in the choice of boundary conditions and (ii)
the problem of thermal description is restricted by quantum
uncertainty principle only.

So, we accommodate the economical thermodynamical description using
the microcanonical approach to work with many-particles system. We
will be seen that this and the canonical Gibbs descriptions (in the
real-time formulation \C{land}) are coincide iff the system is in
equilibrium.  The attempt to extend our thermal description on the
quantum media (nonequilibrium as well) leads to the well known in
statistics Bogolyubov condition on the particles correlation
functions \C{bog}.  Just discussion of this condition allows to show
in a what sense the $S$-matrix approach has the right classical limit.

One should draw attention also on the following factorization property
natural for LSZ reduction formulae (\r{2.1'}). So, the generating
functional may be written as the product:
\be
\R(\b,z)=e^{-N(\b,z;\h\p)}\R_0 (\p).
\l{*'}\ee
The differential over auxiliary field $\p(x)$ operator $N(\b,z;\h\p)$
is the functional of `activity' $z=z(q)$ (to `mark' the momentum $q$ of
external particles) and of temperature
$T=1/\b$ (to define the mean value of the asymptotic states energy).
The physical meaning of this quantities will be discussed.

The functional $\R_0 (\p)$ describes interaction of fields and is
the vacuum-in-vacuum transition $probability$ in presence of
external (auxiliary) field $\p(x)$.  One can say that the operator
$e^{- N(\b,z;\h\p)}$ maps the interacting fields system, described by
$\R_0 (\p)$, on the (external) states marked by given $\b$ and $z$.

This factorization property will have important consequences since it
assumes that external environment do not influence on the spectrum of
quantum excitations. Just in frame of this assumption we would
construct the perturbation theory.

So, at the very beginning we would show that \\

{\it S1. Thermodynamics has the $S$-matrix interpretation.}

The starting point of our calculations is following definition of
$n$-into-$m$ particles transition amplitudes $a_{nm}$ in the momentum
representation \C{elpat}:
\ba
a_{nm} (q';q)
=\prod^{m}_{k=1}\h{\p}(q'_k)\prod^{n}_{k=1}\h{\p}^* (q_k)Z(\p),
\l{2.1'}\ea
where $q'_k$ ($q_k$) are the incoming (outgoing) particles momenta. The
energy-momentum conservation $\d$-function must be extracted from
$a_{nm}$.  The `hat' symbol means variation over corresponding
quantity. For instance, \be \h{\p}(q) \equiv \int dx e^{-iqx}
\f{\d}{\d \p (x)} \equiv \int dx e^{-iqx} \h{\p}(x).  \l{2.2}\ee
$Note$, $\hat{\p}(q)$ acts as the annihilation operator of the
incoming particle and $\h{\p}^* (q)$ as the creation one. The
vacuum-in-vacuum transition amplitude in the background field $\p$ is
\be
Z(\p)=\int Du e^{iS_0 (u) -iV(u+\p)},
\l{2.3'}\ee
where $S_0$ is the free part of action:
\be
S_0 (u)=\f{1}{2}\int_{C_+} dx ((\pa_{\mu}u)^2 - m^2 u^2)
\l{2.4'}\ee
and $V$ describes the interactions:
\be
V(u)=\int_{C_+} dx v(u).
\l{2.5'}\ee
The time integrals in (\ref{2.3'}) are defined on Mills time contour
\C{mil}:
\be
C_+ :t \rar t+i\e,~~~\e \rar +0,~~~-\infty \leq t \leq +\infty.
\l{2.6'}\ee
This guaranties convergence of the path integral. At the end of
calculations one must put the auxiliary field $\p$ equal to zero.

Let us calculate now the probability
\ba
\tilde{\R}_{nm}(P)=\f {1}{n!m!}\int d\o_m (q') d\o_n (q)
\n\\\times
\d (P-\sum^{m}_{k=1}q'_k)\d (P-\sum^{n}_{k=1}q_k)|a_{nm}|^2,
\l{2.7'}\ea
where
$$
 d\o_m (q)=\prod^{m}_{k=1}\f{dq_k}{(2\pi)^3 2\epsilon (q_k)},~
 \epsilon (q)=(q^2 +m^2)^{1/2}
$$
is the Lorentz-invariant phase-space element. $Note$, introduction of
`probability' $\tilde\R$ leads to the doubling of degrees of freedom.

Inserting (\ref{2.1'}) into (\ref{2.7'}) we find:
\ba
\tilde{\R}_{nm}(P)=(-1)^{n+m}\f{1}{n!m!}N_m(P,\h\p)N^*_n(P,\h\p)
\R_0 (\phi),
\l{2.9'}\ea
where
\ba
N_m(P,\h\p)= \int d\o_m (q')\d (P-\sum^{m}_{k=1}q'_k)
\n\\\times
\int
\prod^{m}_{k=1} d x'_k d y'_k e^{-iq'_k (x'_k-y'_k)} \hat{\phi}_-
(y'_k)\hat{\phi}_+(x'_k)
\l{N_m}\ea
and
\be
\R_0 (\phi)=Z(\phi_+)Z^*(-\phi_-).
\l{2.10'}\ee
Introducing new coordinates:
\ba
x_k=R_k+r_k/2,~~~y_k=R_k-r_k/2,~~~
\n\\
x'_k=R'_k-r'_k/2,~~~y'_k=R'_k+r'_k/2
\l{2.11'}\ea
we come naturally from (\ref{2.9'}) to definition of the Wigner
functions \C{carr}:
\ba
\tilde{\R}_{nm}(P)=\f{1}{n!m!}\int d\o_m (q')d\o_n (q)
\n\\\times
\d (P-\sum^{m}_{k=1}q'_k)\d (P-\sum^{n}_{k=1}q_k) \n
\\\times \int \prod^{n}_{k=1}d R_k \prod^{m}_{k=1}d R'_k
W_{mn}(q,R;q',R'),
\l{2.12'}\ea
where
\ba
W_{nm}(q,R;q',R')=(-1)^{n+m}\prod^{n}_{k=1} N_{+}(q_k,R_k; \h\p)
\n\\\times
\prod^{m}_{k=1} N_{-}(q'_k,R'_k;\h \p)
\R_0 (\p),
\l{2.13'}\ea
and
\be
N_{\pm}(R,q; \p)\equiv
\int d r e^{iqr} \hat{\phi}_{\pm}(R+r/2)\hat{\phi}_{\mp}(R-r/2)\}
\l{2.14'}\ee
will be considered as the particles number operator, local in phase
space $(R,q)$.

The Wigner function has formal meaning in quantum case (it is not
positively definite), but in classical limit it has the meaning of
ordinary in statistics phase-space distribution function. It obey the
Liouville equation \C{carr} conserving the phase space volume. We
will use this function searching the classical limit of our
$S$-matrix formalism.

We would consider $d\Gamma_n=|a_{mn}|d\o_n (q)$ as the differential
measure of final state and $d\Gamma_m=|a_{mn}|d\o_m (q')$ of
initial one.  Two $\d$-functions of energy-momentum shells was
introduced separately in (\r{2.7'}) to distinguish the initial and
final states.  The Fourier transform of this $\d$-functions in
(\ref{2.12'}) gives new quantity:
\be
\tilde{\R}_{mn}(P)=
\int\f{d\a_i}{(2\pi)^4}\f{d\a_f}{(2\pi)^4}e^{iP(\a_i+\a_f)}\R_{mn}(\a)
\l{2.16'}\ee
where
\ba
\R_{mn}(\a)= \f{1}{n!m!}\int d\o_m (q') d\o_n (q)
\n \\\times
\prod^{m}_{k=1}d R'_k e^{-i\a_i q'_k}
\prod^{n}_{k=1}d R_k e^{-i\a_f q_k}
W_{mn}(q,R;q',R').
\l{2.17''}\ea

Inserting (\ref{2.13'}) into (\ref{2.17''}) we find:
\be
\R_{mn}(\a)=(-1)^{n+m}\f{1}{n!m!}N_+^m(\a_i;\h\p)N_-^n(\a_f;\h\p)
\R_0 (\phi),
\l{2.18'}\ee
where
\ba
N_{\pm}(\a;\h\p)
\equiv \int dR d\o_1 (q)e^{-iq\a}
\n\\\times
\int d r e^{iqr} \hat{\phi}_{\pm}(R+r/2)\hat{\phi}_{\mp}(R-r/2)\}
\n \\
\equiv \int d R d\o_1 (q)e^{-iq\a}N_{\pm}(R,q;\h\p).
\l{2.19'}\ea

It is natural to introduce the generating functional weighing the
operator $N_{\pm}(R,q;\h\p)$ by the arbitrary `good' function $z(R,q)$:
\be
N_{\pm}(\a,z;\h\p)\equiv \int d R d\o_1 (q)e^{-iq\a}z(R,q)
N_{\pm}(R,q;\h\p).
\l{2.20'}\ee
In result, summation over all $n,m$ gives the generating functional
(\r{*'}):
\ba
\R(\a,z)=e^{-N_+(\a_i,z_i;\h\p)-
N_-(\a_f,z_f;\h\p)}\R_0(\p)
\n\\
\equiv e^{- N(\a,z;\h\p)}\R_0(\p).
\l{2.23'}\ea
It is the generating functional of Wigner functions in the
temperature representation \C{hu}.  At the same time, if $z=z(q)$
then (\r{2.23'}) is the Fourier transform of expectation values
(\r{1.1}) generating functional \C{carr}.  Indeed, in this case
\be
N_{\pm}(\a,z;\h\p)\equiv \int d\o_1 (q)e^{-iq\a}z(q)
\h \p_{\pm}(q)\h \p^*_{\mp}(q).
\l{2.20''}\ee
This representation is suitable for quantum case.

Let us consider the `spectral representation' (\r{2.20''}). It leads
to the following generating functional ($z=1$ is chosen for
simplicity):
\ba
\R(\a)=e^{i\int dx dx'
\hat{\phi}_+(x)G_{+-}(x-x',\alpha_2)\hat{\phi}_-(x')}
\n\\\times
e^{-i\int dx dx'\hat{\phi}_-(x)G_{-+}(x-x',\alpha_1)\hat{\phi}_+(x')}
Z(\phi_+)Z^*(\phi_-),
\l{22'}\ea
where $G_{+-}$ and $G_{-+}$ are  the positive and negative frequency
correlation functions:
\be
G_{+-}(x-x',\alpha)=-i\int d\omega(q)e^{iq(x-x'-\alpha)}
\l{23''}
\ee
describes the process of particles creation at the  time  moment $x_0$
and its absorption at $x'_0$, $x_0>x'_0$, and $\a$ is the
center of mass (CM) 4-coordinate. The function
\be
G_{-+}(x-x',\alpha)=i\int d\omega(q)e^{-iq(x-x'+\alpha)}
\l{24'}
\ee
describes the opposite process, $x_0<x'_0$. These functions obey the
homogeneous equations:
\be
(\partial^2 +m^2)_x G_{+-}=
(\partial^2 +m^2)_x G_{-+}=0
\l{25'}
\ee
since the `propagation' of mass-shell particles is described.

Let us suppose that $Z(\p)$ may be computed perturbatively. Following
transformation, suitable for the arbitrary nonsingular at origin
functional, would be useful:
\ba
e^{-iV(\phi)}= e^{-i\int dx \hat{j}(x)\hat{\phi}'(x)}
e^{i\int dx j(x)\phi (x)} e^{-iV(\phi ')}=
\n\\ =
e^{\int dx \phi(x)\hat{\phi}'(x)} e^{-iV(\phi ')}= \n\\
=e^{-iV(-i\hat{j})} e^{i\int dx j(x)\phi (x)},
\l{26'} \ea
where $\hat j$ and $\h \p$ are variational derivatives over
corresponding quantities.  At the end of calculations the auxiliary
variables $j$, $\p'$ should be taken equal to zero.

Using first equality in (\ref{26'}) we find that
\ba
Z(\phi)= e^{-i\int dx \hat{j}(x)\hat{u}(x)} e^{-iV(u+\phi)}
\n\\\times
e^{-\frac{i}{2}\int dx dx' j(x)G_{++}(x-x')j(x')},
\l{27'} \ea
where
$G_{++}$ is the causal Green function:
\be
(\pa^2 +m^2)_x G_{++} (x-y)=\delta (x-y) .
\l{28'} \ee

Inserting (\ref{27'}) into (\ref{22'}) after simple manipulations
with differential operators, see (\ref{26'}), we find the expression:
\ba
\R(\a)= e^{-iV(-i\hat{j}_+)+iV(-i\hat{j}_-)}
\n\\\times
e^{ \frac{i}{2} \int dx dx'
 j_i (x)G_{ik}(x-x')j_k (x')},~i,k=1,2
\l{29'}\ea
where
\ba
G_{11}(x-x')=-G_{++}(x-x'),
\n\\
G_{12}(x-x')=G_{+-}(x-x',\a_1),
\n \\
G_{21}(x-x')=-G_{-+}(x-x',\a_2),
\n\\
G_{22}=G_{--}=(G_{++})^*,
\l{30'}\ea
$G_{--}$ is the anticausal Green function.

The structure of generating functional (\ref{29'}) is the same as in
the real-time finite-temperature field theories, e.g. \C {sem}.
The difference is only in definition of Green functions.

The Green functions $G_{ij}$ were defined on the time contours
$C_{\pm}$ in the complex time plane ($C_-=C_+^*$). This definition of
the  time contours coincide with Keldysh' time contour \C{kel}. The
expression (\ref{29'}) was written in the compact matrix form \C{moh}.

Note, the doubling of degrees of freedom is $unavoidable$ since
Green functions $G_{ij}$ are singular on the light cone. But it will
be seen below that one can shift the time contour on the real axis if
the perturbation theory is constructed in the invariant subspaces.

Considering the system with large number of particles we can simplify
calculations choosing the CM frame $P=(P_0 =E,\vec 0)$.  It is useful
also \C{kaj} to rotate the contours of integration over
$\alpha_0$: $\alpha_0=-i\b,~Im\b =0$. In result, omitting unnecessary
constant, we will consider $\R=\R(\b,z)$. Note, $\b$ is conjugate to
particles energy, i.e. $1/\b$ has the meaning of temperature.

So, we construct the two-temperature theory (for initial and final
states separately). In such theory with two temperatures
the Kubo-Martin-Schwinger (KMS) \C{sch, kubo} periodic boundary
conditions applicability is not evident. $Note$, KMS boundary
condition play the crucial role in Gibbs thermodynamics since the
temperature in it is introduced just by this condition, e.g. \C{sem},
see (\r{kms}).

Let us consider the dynamical origin of KMS condition, see also
\C{elpat}.  By definition, the path integrals
\ba
\R_0 (\phi_{\pm})=\int Du_+ Du_- e^{iS_0(u_+)-iS_0(u_-)}
\n\\\times
e^{-iV(u_++\phi_+) + iV(u_- -\phi_-)},
\l{35'}\ea
should describe the {\it closed path motion} in the space of fields
$u$. The equality:
\be
\int_{\sigma_{\infty}}
d\sigma_{\mu} u_+ \partial^{\mu}u_+ = \int_{\sigma_{\infty}}
d\sigma_{\mu} u_- \partial^{\mu}u_-.
\l{36'} \ee
takes into account this boundary condition. So, $\R (\b,z)$ is
defined on the periodic (in the $u$ space) trajectories by
definition.

Mostly general solution of eq.(\ref{36'}) means that the fields
$u_+$ and $u_-$ (and theirs first derivatives $\pa_{\mu}u_{\pm}$)
must coincide on the boundary hypersurface $\s_{\infty}$:
\be
u_{\pm}(\sigma_{\infty})=u(\sigma_{\infty}),
\l{37'}\ee
where, by definition, $u(\sigma_{\infty})$ is an arbitrary,
`turning-point', field. The value of $u(\s_{\infty})$ specify the
environment of the system. This boundary condition guaranties absence
of surface terms up to `non-integrable' term \C{yad}. Last one can
arise if the topology of interacting fields is nontrivial (see e.g.
Sec.5).

The simplest (minimal) choice of $u(\s_{\infty})\neq 0$ assumes
that the  system under consideration is surrounded by black-body
radiation. One should underline also that this choice of boundary
condition is not unique: one can consider another organization of the
environment, e.g. the external flow can be consist of the correlated
particles as it happens in the heavy ion collisions (the nucleons in
ion may be considered as the quasi-free particles).

Let us suppose that on the infinitely far hypersurface $\s_{\infty}$
there are only free, mass-shell, particles. This assumption is
natural in the $S$-matrix framework \C{lan}.  In this paper we will
assume also that there are not any special correlations among
background particles.

In this framework our derivation is the same as in \C{psf}. By this
reason we restrict ourselves mentioning only the main quantitative
points.

Calculating the  product $a_{n,m}a^*_{n,m}$ we describe a process of
particles creation and theirs further  adsorption. In the vacuum case
two {\it time ordered process} of particles creation and absorption
were taken into account. In presence of the background particles this
time-ordered picture is slurring over since the possibility to absorb
particles before its creation appears. Taking new possibilities
into account,
\be
\R_{cp}=e^{iN(\b,z;\hat{\phi})}\R_0(\p),
\l{44'}\ee
where $\R_0 (\p)$ is the same generating functional, see
(\r{35'}).

The operator $N(\b,z;\hat{\p}), i,j=+,-,$
describes the external particles environment. It can be expanded over
the activity operator $\hat\p^*_i(q)\hat\p_j(q)$. We can leave only
the first nontrivial term:
\be
N(\b,z;\hat{\phi})=
\int d\omega_1 (q) \hat{\phi}^*_i (q) n_{ij}(\b,z) \hat{\phi}_j (q),
\l{45'}\ee
since no special correlation among background particles should be
expected. Following to our interpretation of $\hat\p^*_i\hat\p_j$ we
conclude that $n_{ij}$ is the mean multiplicity (occupation number)
of background particles. In (\ref{45'}) the  normalization condition:
\be
N(0)=0
\l{46'}
\ee
was used and summation over all $i,j$ was assumed. In the vacuum case
only the combinations $i\neq j$ are present.

Having background particles flow it is important to note that to each
vertex of in-going in $a_{n,m}$ particle we must adjust the factor
$e^{-i\a_1q/2}$ and for each out-going particle we have
correspondingly $e^{-i\a_2q/2}$, see (\r{2.17''}).

So, the product $e^{-i\a_kq/2}e^{-i\a_jq/2}$ can be interpreted as
the probability factor of the one-particle $(creation+annihilation)$
process. The $n$-particles $(creation+annihilation)$ process'
probability is the simple product of  these factors if  there is not
the special correlations among background  particles. This
interpretation is evident in the CM frame $\a_k=(-i\b_k,\vec0)$.

After this preliminaries it is not too hard to find $\bar{n}_{ij}$
(see Appendix A). Corresponding generating functional has the
standard form:
\ba
\R_{p}(j_{\pm})=\exp\{-iV(-i\hat{j}_+)+iV(-i\hat{j}_-)\}\times
\n \\ \times
\exp\{\frac{i}{2}\int dx dx' j_i (x)G_{ij}(x-x',\b)j_j(x')\}
\l{53'}
\ea
where the summation over repeated indexes  is assumed.

Inserting (\ref {53'}) in the equation of state (\ref{es}) we can find
that $\beta_1 =\beta_2 =\beta (E)$. If $\beta (E)$ is a  `good'
parameter then $G_{ij}(x-x';\beta )$ coincide with the Green
functions of the real-time finite-temperature field theory and the
KMS boundary condition:
\ba
G_{+-}(t-t')=G_{-+}(t-t'-i\beta),\;\;\;
\n\\
G_{-+}(t-t')=G_{+-}(t-t'+i\beta),
\l{54'}\ea
is restored. The eq.(\ref{54'}) deduced from (\ref{53'}) by direct
calculations. It is known that the KMS boundary condition without
fail leads to the $equilibrium$ fluctuation-dissipation conditions
\C{haag} (see also \C {chu}).

The energy and momentum in our approach are still locally conserved
quantities since an amplitude $a_{nm}$ is translational invariant.
So, we  can perform evident in the $S$-matrix theory transformation:
\ba
\a_1 \sum q_k =(\a_1 -\sigma_1 )\sum q_k
+\sigma_1 \sum q_k \rar
\n\\
(\a_1 -\sigma_1 )\sum q_k +\sigma_1 P
\l{'''}\ea
since 4-momenta are conserved. The choice of $\sigma_1$ defines the
reference frame. This degree of freedom of the theory was considered
in \C{mta,psf}. It gives the rule as the time contour can be shifted.

Using this rule it is not hard to present $\R(\b)$ in the form of one
path integral defined on the `closed time-path' contour $C$ \C{sem}:
\be
\R(\b)=\int Du e^{iS_C (u)},
\l{sem}\ee
where the action $S_C$ is defined on the $Mills$ \C{mil} time contour
$C$. It start at time $t_i$ goes to right, at $t=t_f$ it turns and
end at $t_i -i\b$.  The temperature is introduced through the KMS
boundary condition
\be
u(t_i)=u(t_i -i\b).
\l{kms}\ee
$Note$, one can find (\r{53'}) from (\r{sem}) considering $t_i \rar
-\infty$ and $t_f \rar +\infty$ iff the interactions are
disappeared on $\s_ \infty$.  In the ordinary perturbation theory
this condition is hold \C{land}.  But the symmetries (hidden as well)
may survive the `interactions'.

The contour $C$ unavoidably contains both along the real and
imaginary axis parts. It is not clear by this reason how the
nontrivial topology requirements can be applied for such time
contour. In contrast with it the representation (\r{53'}) is free
from the imaginary parts of time contour.\\

$\bullet$ The temperature was introduced as the parameter conjugate to
created particles energies. By this reason the uncertainty principle
restrictions should be taken into account. We would like to show that\\

{\it S2. The thermal $S$-matrix description can be used for infrared
stable field theory.}

Considering the Fourier-transformed probability $\R(\a,z)$ as the
observable quantity the phase-space boundaries are not fixed exactly,
i.e. the 4-vector $P$ can be defined with some accuracy only if
$\a_i$ are fixed, and vice versa. It is the ordinary quantum
uncertainty condition. In the particles physics namely the 4-vector
$P$ is defined by experiment. Let us find the condition when both $P$
and $\b$ may be the well defined quantities, i.e. may be used for
description of $\R(\a,z)$.  This is necessary if we want to use the
temperature formalism in particles physics also.

Note, in statistical physics such formulation of problem has no
meaning since the interaction with thermostat is assumed. In result
of this interaction the energy of system is not conserved, i.e. the
systems word line belong to the energy surface\footnote{It can be
thin if the interaction with thermostat is weak.}.

The stationary phase condition for integrals (\r{2.16'}) gives the
equations of (final and initial) states familiar for microcanonical
description \C{elpat}. We will chose the CM frame $P=(E, \vec 0)$
when $\a_{i(f)}=(-i\b_{i(f)},\vec{0})$. We can interpret
$1/\b_{i(f)}$ as the temperature of initial (final) state. The
corresponding equations of state have the form:
\be
E=-\f{\pa}{\pa\b_{i(f)}}\ln \R (\b,z).
\l{es}\ee
But one can not define $\tilde{\R}(E)$ correctly even knowing the
solutions $\b_{i(f)}(E)$ of eqs.(\r{es}) if $\b_{i(f)} (E)$ are not a
`good' parameters, i.e. iff the fluctuations in a vicinity of
them are not Gaussian. This condition would be considered as the
definition of equilibrium.

Indeed, to calculate the integrals over $\b_{i(f)}$ the expansion near
solutions $\b_{i(f)}(E)$ of eqs.(\r{es}) should be examined.
This leads to asymptotic series with coefficients
\ba
\sim\int \prod_i^s d\o_1 (q_i) <\e (q_1), \e (q_2),..., \e (q_s)>
\equiv\n\\
\int {\bf D}_s (q_1,q_2,...,q_s),
\ea
since $\ln\R (\b ,z)$ is the essentially nonlinear function of $\b$.
So, the fluctuations near $\b_{i(f)}(E)$ are defined by the value of
$s$-particle inclusive energy spectra $<\e (q_1), \e (q_2),..., \e
(q_s)>$ familiar in particles physics. The analysis shows that it is
enough to have the factorization \C{elpat}:  \ba \int \prod^s d\o
(q_i) <\e (q_1),\e (q_2),...,\e (q_s)> \sim \n\\\prod^s \int d\o (q_i)
<\e (q_i)> \l{rel}\ea for correct estimation of this asymptotic
series.  It must be underlined that this is the unique solution of
the thermal descriptions problem.

Discussed factorization is the well known Bogolyubov's condition
for `truncated' description of nonequilibrium media \C{bog}, when $s$-
particle distribution functions ${\bf D}_s$, $s>1$, is expressed through
${\bf D}_1$ through the relation (\r{rel}).

Considering the general problem of particles creation it is hard to
expect that the constant $\b_{i(f)}(E)$ is a `good' parameter, i.e. that
the factorization conditions (\r{rel}) are hold. Nevertheless there
is a possibility to have the above factorization property in the
restricted space-time domains of size $L$. It is the so called
`kinetic' phase of the process when the memory of initial state was
disappeared, the `fast' fluctuations was averaged over and we can
consider the long- range `slow' fluctuations only.

In this `kinetic' phase one can use the `local equilibrium'
hypothesis in frame of which $\b_{i(f)}(E) \rar \b_{i(f)}(R,E)$,
where $R \in L_c$ and $L_c$ is the dimension of the $measurement$
cell\footnote{Note, we always can divide the external particles
measuring device on cells since the in free state are measured. Such
description of nonstationary media seems favorable in comparison
with traditional one, e.g. \C{zub}}.  It is natural to take
\be
L_c << L,
\l{con1}\ee
where $L$ is the characteristic thermal fluctuations dimension. It is
assumed that $\b_{i(f)}(R,E) =const.$ if $R \in L_c$. In the
equilibrium $L\rar\infty$.

By definition, $1/\b_{i(f)}(R,E)= \bar{\e}(R)$ is the mean energy of
particles in the cell with dimension $L_c$. The fluctuations in
$\bar{\e}(R)$ vicinity should be Gaussian.

The quantum uncertainty principle dictates also the condition \C{lan}:
\be
L_c >> L_q,
\l{con2}\ee
where $L_q$ is the characteristic scale of quantum fluctuations ($L_q
\sim 1/m$ for massive theories).

The `infrared unstable' situation means that
\be
L_q >> L.
\l{inf}\ee
One should underline that $L$ defines the scale of thermodynamics
fluctuations and, by this reason, the inequality (\r{inf}) points to
(unphysical) instability in the infrared domain.

So, if conditions (\r{con1}, \r{con2}) are hold we can use the Wigner
functions to describe the phase-space distributions, i.e. the
formalism has right classical limit in this case.

To introduce the scales $L,~L_c$ into formalism we can divide the $R$
4-space on the cells of $L_c$ dimension \C{elpat}:
\be
\int dR=\sum_{r}\int_{L_c}dR,
\l{cel}\ee
where $r$ can be considered as the cells 4-coordinate. Assuming
that the inequality (\r{con1}) is hold we can assume that
$
\b=\b(r),~~z=z(q,r),
$
are the constants at least on the $L_c$ scale. With this definitions
\ba
\h N_{\pm}(\b,z;\p)\equiv \sum_r \int d\o_1 (q)e^{-\b(r)(\e(q)+\mu(r,q))}
\n\\\times
\int_{L_c} d R\h N_{\pm}(R,q;\p),
\l{Lc}\ea
where
$$
\mu(r,q)\equiv\f{1}{\b(r)}\ln z(r,q)
$$
is the local chemical potential.\\

$bullet$ On the more early pre-kinetic stages no thermodynamical
shortened description can be applied and the pure quantum description
(in terms of momenta only) should be used. For this purpose one
should expand $\R(\a,z)$ over operators $ N_\pm$ and the integrations
over $\a_i,~\a_f$ gives ordinary energy-momentum conservation
$\d$-functions, i.e.  defines the system on the infinitely thin
energy sheet.

\section{Unitarity condition}\0

Purpose of this section is to show how the $S$-matrix unitarity
condition can be introduced into the path-integral formalism to find
measure (\r{a'}) \C{yad}. We will start from the quantum-mechanical
example to do the calculations more evident.

The unitarity condition for the  $S$-matrix $SS^+ =I$ presents the
infinite set of nonlinear equalities:
\be iA A^* =A - A^*,
\l{2'}\ee
where $A$ is the amplitude, $S=I+iA$\footnote{$Note$, in this
definition the amplitude is dimensionless. In ordinary definitions
the energy-momentum conservation $\d$-functions are extracted from
amplitudes.}.  Expressing the amplitude by the path integral one can
see that the l.h.s. of this equality offers the double integral and,
at the same time, the r.h.s.  is the linear combination of integrals.
Let us consider what this linearization of product $AA^*$ gives.

Using the spectral representation of one-particle amplitude:
\be
A_1(x_1 ,x_2 ;E)=\sum_{n}\frac{\Psi^*_{n} (x_2)\Psi_n (x_1)}{E-E_n
+i\e}, ~~~\e \rar +0,
\l{3'}\ee
let us calculate
\be
\R_1(E)=\int dx_1 dx_2 A_1(x_1 ,x_2 ;E)A^*_1 (x_1 ,x_2 ;E).
\l{4'} \ee
The integration over end points $x_1$ and  $x_2$ is performed for
sake of simplicity only. Using ortho-normalizability of the wave
functions $\Psi_{n} (x)$ we find that
\be
\R_1(E)=\sum_{n}|\frac{1}{E-E_n+i\e}|^2 =\frac{\pi}{\e}\sum_{n}\d
(E-E_{n}).
\l{5'}\ee
Certainly, the last equality is nothing new but it is important to
note that $\R_1(E) \equiv 0$ for all $E\neq E_n$, i.e. that all
unnecessary contributions with $E\neq E_n$ were canceled by
difference in the r.h.s. of eq.(\ref{2'}).  We will put this
phenomena in the basis of the approach.

We will build the perturbation theory for $R(E)$ using the
path-integral definition of amplitudes \C{yad}. It leads to loss of
some information since the amplitudes can be restored in such
formulation with the phase accuracy only. Yet, it is sufficient for
calculation of the energy spectrum.  We would consider this quantity
to demonstrate following statement:\\

{\it S3. The unitarity condition unambiguously determines
contributions in the  path integrals for $\R_1(E)$.}

This statement looks like a tautology since $e^{iS(x)}$, where
$S(x)$ is the quantum-mechanical action, is the unitary operator
which shifts a system along the trajectory\footnote{It is well known
that this unitary transformation is the analogy of tangent
transformations of classical mechanics \C{fok}.}. I.e. the unitarity
is already included in the path integrals. But the general
path-integral solution contains unnecessary degrees of freedom
(unobservable states with $E\neq E_n$ in our example).  We would
define the quantum measure $DM$ in such a way that the condition of
absence of unnecessary contributions in the final (measurable) result
be loaded from the very beginning. Just in this sense the unitarity
looks like the necessary and sufficient condition unambiguously
determining the complete set of contributions. Solution is simple:
one should find, as it follows from (\r{5'}), the linear
path-integral representation for $\R_1(E)$ to introduce this condition
into the formalism.

Indeed, to see the integral form of our approach, let us use the
proper-time representation:
\be
A_1(x_1 ,x_2 ;E)=\sum_{n} \Psi_{n}
(x_1)\Psi^{*}_{n} (x_2)i
\int^{\infty}_{0}dTe^{i(E-E_{n}+i\e)T}
\l{6'} \ee
and insert it into (\ref{4'}):
\be
\R_1(E)=\sum_{n} \int^{\infty}_{0} dT_{+}dT_{-}
e^{-(T_{+}+T_{-})\e} e^{i(E-E_{n})(T_{+}-T_{-})}.
\l{7'} \ee

We  will introduce new time variables instead of $T_{\pm}$:
\be T_{\pm}=T\pm\tau,
\l{8'} \ee
where, as follows from Jacobian of transformation, $|\tau|\leq
T,~0\leq T\leq \infty$. But we can put $|\tau|\leq\infty$ since
$T\sim1/\e\rar\infty$ is essential in integral over
$T$. In result,
\be
\R_1(E)=2\pi\sum_{n}\int^{\infty}_{0} dT
e^{-2\e T} \int^{+\infty}_{-\infty}\f{d\tau}{\pi}
e^{2i(E-E_{n})\tau}.
\l{9'} \ee
In the last integral all contributions with $E\neq E_{n}$ are
canceled.  Note that the product of amplitudes $AA^*$ was
`linearized' after introduction of `virtual' time $\tau
=(T_{+}-T_{-})/2$. The physical meaning of such variables will be
discussed, see also \C{fok}.

We will consider following path-integral:
\be
A_1(x_1 ,x_2;E)=i\int^{\infty}_{0}dT
e^{iET}\int_{x_1=x(0)}^{x_2=x(T)} Dx e^{iS_{C_+}(x)},
\l{3.1}\ee
where $C_+ $ is the Mills complex time contour \C{mil}. Calculating
the probability to find a particle with energy $E$ ($Im~E$ will not
be mentioned for sake of simplicity) we have:
\ba
\R_1(E)=\int dx_1 dx_2 |A|^2 =
\int^{\infty}_{0} dT_+ dT_- e^{iE(T_+-T_-)}
\n\\\times
\int_{x_+ (0)=x_-(0)}^{x_+ (T_+)=x_-(T_-)} D_{C_+}x_+
D_{C_-}x_-
\n\\\times
e^{iS_{C_+ (T_+ )}(x_+ ) -iS_{C_- (T_- )}(x_- )},
\l{3.2a}\ea
where $C_- (T)=C^{*}_{+}(T)$. Note that the total action in
(\r{3.2a}) $(S_{C_+ (T_+ )}(x_+ ) - S_{C_- (T_- )}(x_-))$ describes
the closed-path motion by definition.

New independent time variables $T$ and $\tau$ will be used again, see
(\r{8'}). We will introduce also the mean trajectory $x(t)=(x_+(t)
+x_-(t))/2$ and the deviation $e(t)$ from it: $x_{\pm}(t)=x(t)\pm
e(t)$. Note that one can do surely this linear transformations in the
path integrals.

We will consider $e(t)$ and $\tau$ as the fluctuating, virtual,
quantities and calculate the integrals over them perturbatively. In
the zero order over $e$ and $\tau$, i.e. in the quasiclassical
approximation, $x$ is the classical path and $T$ is the total time of
classical motion.

The boundary conditions (see (\r{3.2a})) states the closed-path
motion and therefore we have the boundary conditions for $e(t)$ only:
\be
e(0)=e(T)=0.
\l{3.6}\ee
Note the uniqueness of this solution if the integral over $\tau$ is
calculated perturbatively.

Extracting the linear over $e$ and $\tau$ terms from the closed-path
action ($S_{C_+ (T_+ )}(x_+ ) - S_{C_- (T_- )}(x_- )$) and expanding
over $e$ and $\tau$ the remainder terms:
\be
-\tilde {H}_T (x;\tau)=(S_{C_+ (T+\tau)}(x) - S_{C_- (T-\tau )}(x))+
2\tau H_T (x),
\l{3.7}\ee
where $H_T(x)$ is the Hamiltonian at the time moment $T$, and
\ba
-U_T (x,e)=(S_{C_+ (T)}(x+e)-S_{C_- (T)}(x-e))+
\n\\
2\Re\int_{C_+(T)}dt(\ddot{x}+ v'(x))e
\l{3.8}\ea
we find that
\ba
\R_1(E)=2\pi \int^{\infty}_{0}dTe^{-i\h{K}(\o ,\tau;j,e)}
\n\\\times
\int DM(x)e^{-i\tilde{H}_T (x;\tau)-iU_T (x,e)}.
\l{3.10}\ea
Note the necessity of boundary condition (\r{3.6}) to find
(\r{3.10}). It allows to split the expansions over $\tau$ and $e$.

The  expansion over differential operators:
\be
\h{K}(\o ,\tau;j,e)=
\frac{1}{2}(\frac{\partial}{\partial\o}\frac{\partial}{\partial \tau}
+\Re\int_{C_+(T)}dt\frac{\delta}{\delta j(t)}\frac{\delta}{\d e(t)})
\l{3.11}\ee
will generate the perturbation series. We will assume that it exist at
least in Borel sense.

In (\r{3.10}) the functional measure
\be
DM(x)=\delta (E+\o -H_T(x))\prod_t dx(t) \delta (\ddot{x}+v'(x)-j)
\l{3.12}\ee
unambiguously defines the complete set of contributions in the path
integral. The functional $\delta$-function is defined as follows:
\ba
\prod_t
\delta (\ddot{x}+v'(x)-j)
\n\\
=(2\pi )^2 \int_{e(0)}^{e(T)}
\prod_{t}\frac{de(t)}{\pi} e^{-2i\Re\int_{C_+}dt
e(\ddot{x}+v'(x)-j)} \l{*}\ea

The physical meaning of this $\d$-function is following. We can
consider $(\ddot{x}+v'(x)-j)$ as the total force and $e(t)$ as the
virtual deviation from true trajectory $x(t)$. In classical mechanics
the virtual work must be equal to zero: $(\ddot{x}+v'(x)-j)e(t)=0$
(d'Alembert) \C{arn} since the motion is time reversible. From this
evident dynamical principle one can find the `classical' equation of
motion:
\be
\ddot{x}+v'(x)=j,
\l{3.13}\ee
since $e(t)$ is arbitrary.

In quantum theories the virtual work usually is not equal to zero,
i.e.  the quantum motion is not time reversible since the quantum
corrections can shift the energy levels. But integration over $e(t)$,
with boundary conditions (\ref{3.6}), leads to the same result. So,
in quantum theories the unitarity condition \C{fok} play the
same role as the d'Alembert's variational principle in classical
mechanics. We can conclude, the unitarity condition as the dynamical
principle establish the $time-local$ equilibrium between classical
(r.h.s. of (\r{3.13})) and quantum (l.h.s. of (\r{3.13})) forces.\\

$\bullet$ We would like to show that \\

{\it S4. The functional measure of $\R(\b ,z)$ is $\d$-like.}

Let us consider now the integrals
\ba
\R_0(\phi)=\int Du_+ Du_- e^{iS_0 (u_+) -iU(u_+ +\phi_+)}
\n\\\times
e^{-iS_0 (u_-) +iV(u_- -\phi_-)},
\l{3.1'}\ea
where $S_0$, $V$ were defined in (\ref{2.4}), (\ref{2.5}) and
$(u_-,~\phi_-)$ are defined on the complex conjugate time contour
$C_-$.
$Note$, the fields $\p_\pm$ carry all external information and
this integrals should be calculated with `closed-path' boundary
condition, see (\r{36'}).

Instead of two independent fields $u_+$ and $u_-$ we will use new
ones \C{yad}:
\be
u(x)_{\pm}=u(x)\pm\vp(x)
\l{3.3}\ee
with `closed-path' boundary condition:
\be
\int_{\s_{\infty}} dx_{\mu}\vp(x)\pa^{\mu}u(x)=0,
\l{3.4}\ee
where $\s_{\infty}$ is the remote hypersurface. We will choose
following solution of (\ref{3.4}):
\be
\vp(x)|_{(x)\in \s_{\infty}}=0
\l{3.4'}\ee
With this boundary condition the total action $(S_0 (u_+) -V(u_+)
-S_0 (u_-) +V(u_-))$ describes the closed-path motion with
turning-point field $u(x)|_{x\in \s_{\infty}}$. The integration over
it is assumed.  The physical meaning of this `minimal' boundary
condition in the $S$-matrix approach was described in Sec.3
\C{elpat}.

We will consider $\vp$ as a pure quantum field expanding (\ref{3.1'})
over them. Introducing the auxiliary field $\p (x,t)$:
$$
\p(x,t\in C_{\pm})=\phi_{\pm} (x,t \in C_{\pm})
$$
and introducing the variational derivative by equality:
$$
\f{\d \phi(x,t\in C_i)}{\d \phi(x',t'\in C_j)}=\d_{ij}\d (x-x') \d
(t-t'),~~~ i,j=+,-.
$$
we can write $(Im~\a_{i(f)}<0)$:
\ba
N(\a,z;\h\p)_{\pm}=\int d\o (q)\int_{C_+}d x \int_{C_-}d y
\hat{\phi}(x)\hat{\phi}(y)
\n\\\times
e^{-iq\a_{i(f)}}z_{i(f)} e^{\pm iq(x-y)}.
\l{3.6'}\ea
Note, introducing the Wigner coordinates, see (\r{2.11'}),
$R=(x+y)/2$ and $r=x-y$ we find taking into account (\r{3.6'}) that
$\Im R=0$ and $\Im r\sim\e$ since $C_-=C_+^*$. Therefore, $\a_{i(f)}
(R)$ and $z_{i(f)}(q,R)$ stay real on the complex time contours.

Using this notations let us extract in the exponents (\ref{3.1'}) the
linear over $(\phi +\vp)$ term:
\ba
V(u+(\p+\vp)) - V(u- (\p+\vp))=
U(u,\p+\vp)
\n\\+2\Re\int_{C_+} dx (\phi (x) +\vp (x))v'(u),
\l{3.7'}\ea
and
\ba
S_0 (u+ \vp) -S_0 (u- \vp)=s_0 (u)
\n\\
 -2i\Re\int_{C_+} dx\vp (x) (\pa_{\mu}^2 +m^2 )u(x).
\l{s_0}\ea
where
$$
2\Re\int_{C_+}=\int_{C_+} + \int_{C_-}.
$$
$Note$, generally speaking, $s_0 (u)\neq 0$ if the topology of field
$u(x)$ is nontrivial. The reason of this phenomena was demonstrated
in \C{yaph} on the quantum-mechanical examples. It has the meaning of
`nonintegrable' term.

The expansion over $(\p+\vp)$ can be written in the form, see
(\r{26'}):
\ba
e^{-iU(u,\phi+\vp)}=e^{\f{1}{2i}Re\int_{C_+}dx \hat{j}(x)
\hat{\vp'}(x)}
\n\\\times
e^{i2\Re\int_{C_+}dx dt j(x)(\phi(x)+\vp(x))}e^{-iU(u,\vp')},
\l{3.9}\ea
where $\h{j}(x),~\h{\vp'}(x)$ are the variational
derivatives. The auxiliary variables $(j,\vp')$ must be taken equal
to zero at the very end of calculations.

In result,
\ba
\R_0(\phi)=e^{\f{1}{2i}\Re\int_{C_+} dx \hat{j}(x) \hat{\vp}(x)}\int
Du e^{is_0(u)}e^{-iU(u,\vp)}
\n\\\times
e^{i2\Re\int_{C_+}dx (j(x)-v'(u)) \phi(x)}
\n\\\times
\prod_{x}\d (\pa_{\mu}^2 u + m^2 u +v'(u) -j),
\l{3.15k}\ea
where the functional $\d$-function was defined by the equality:
\ba
\prod_{x}\d (\pa_{\mu}^2 u + m^2 u +v'(u) -j)
\n\\
=\int D\vp e^{-2i\Re\int_{C_+}dx (\pa_{\mu}^2 u + m^2 u +v'(u) -
j)\vp(x)} \l{3.16}\ea The eq.(\r{3.15k}) can be rewritten in the
equivalent form:  \ba \R_0(\phi)=e^{-i\hat{K}(j,\vp)}\int
DM(u)e^{is_0 (u) -iU(u,\vp)}\n\\\times e^{i2Re\int_{C_+}dx \phi
(x)(\pa_{\mu}^2 + m^2) u(x)} \l{3.38}\ea because of the
$\d$-functional measure \be DM(u)=\prod_{x} du(x) \d (\pa_{\mu}^2 u +
m^2 u +v'(u)  -j), \l{3.40}\ee
with
\be
\h{K}(j,\vp)=\f{1}{2}\Re\int_{C_+} dx \hat{j}(x) \hat{\vp}(x).
\l{3.39}\ee
Not at the end that the contour $C_+$ in (\r{3.6'}) and (\r{3.39})
can not be shifted on the real time axis.\\

$\bullet$ It is easy to show now having definition (\r{3.40}) that\\

{\it S5. The expectation values $\R_{nm}$ has the (\r{1.1}) form iff
the field theory is infrared stable.}

The action of operator $N(\b,z;\h\p)$ maps the interacting fields
system on the physical states. Last ones are `marked' by $z_{i(f)}$
and $\b_{i(f)}$. The operator exponent is the linear functional
over $\p$ and this allows easily find the result of mapping:
\be
\R(\b,z)=e^{-i\h{K}}\int DM(u)e^{+is_0(u)-iU(u;\vp)}e^{N(\b,z;u)},
\l{3.41}\ee
where
$$
N(\b,z;u)=n(\b_i,z_i;u)+n^*(\b_f,z_f;u)
$$
and
\ba
n(\b,z;\p)=\sum_{r}\int d\o (q) dk \d_{L_c} (k)
e^{-\b (r)(\e (q) +\mu (q,r))}
\n\\\times
\Ga(q+k,\p) \Ga^* (q-k,\p).
\l{d}\ea
if the thermodynamical parameters $\b$ and $z$ are local quantities.
In (\r{d})
$$
\d_{L_c} (k) =\int_{L_c}\f{dR}{(2\pi)^{4}} e^{ikR}
$$
was introduced (for $4$-dimensional theory). Here $L_c$ is the
space-time dimension where $\b_{i(f)}(R)$ and $z_{i(f)}(q,R)$ can be
considered as the constants. If $L_c << L$ then
$\d_{L_c} (k)$ can be replaced on the usual $\d$-function $\d (k)$
and, therefore, in this limit:
\be
N(\b,z;u)=\int  d\o (q) dr \s(r)
e^{-\b (r)(\e(q)+\mu (q,r))} |\Ga (q,u)|^2
\l{3.42}\ee
Integration over $r$ means summation over cell
coordinates, where the factor $\s (r)$ is the measure of
this replacement. The translational invariantness gives $\s (r)=1$.In
this expression \be \Ga (q,u)=\int_{C_+}dx e^{iqx} (\pa^2 +m^2)u(x)
\l{3.43}\ee
is the function of external particles momentum $q$ only.

The considered limit is hold if the theory is `infrared stable'. In
opposite case the eq.(\r{d}) must be used because of arbitrary-range
quantum fluctuations. Note, in this case the quantum description
is not hold also. This is typical unnphysical instability which may
arise if the ground state is unstable.\\

$\bullet$ Note that
\be
\sum_{nm}\int dP \tilde{\R}_{nm}(P) =\D A_0,
\l{abs}\ee
where $\tilde{\R}_{nm}(P)$ was defined in (\r{2.7'}), is the
absorption part of vacuum-to-vacuum amplitude $A_0=<vac|vac>$.
We found in previous section the functional measure for $\D A_0$
using the unitarity condition. It was enough to know this quantity to
reconstruct the real-time finite-temperature field theory which is
the analytic continuation of the ordinary in statistics Matsubara
approach \C{mats}. It was shown also the way as our microcanonical
approach can be used for nonequilibrium media description and the
condition when our formalism has the right classical limit. For this
purpose the Wigner functions was used.

At the end, without evident calculations, we would like to note that\\

{\it S6. The absorption part of the elastic amplitude $A_2(q_1, q_2;
q_1', q_2')$ is defined by the expression:}
\ba
\D A_2(q_1, q_2; q_1', q_2';\a_f,z_f)
\n\\
=\prod_{i=1}^2
\h{\p}(q_i)\h{\p}^*(q_i')e^{-N(\a_f,z_f;\h{\p})}e^{-i\hat{K}(j,\vp)}
\n\\\times
\int DM(u)e^{is_0 (u) -iU(u,\vp)}
e^{i2\Re\int_{C_+}dx \phi (x)(\pa_{\mu}^2  + m^2) u(x)},
\l{elamp}\ea

We leave in this expression the $(\a,z)$ dependence to show as the
observables can be described.

This formulae is important since having $\D A_2$ and using
dispersion relations one can find the total amplitude $A_2(q_1, q_2;
q_1', q_2')$ which is the main quantity in particles physics.

\section{Perturbation theory}\0

Now let us consider representation (\r{3.10}). It is not hard to show
that\\

{\it S7. Eq.(\r{3.10}) restores the perturbation theory of stationary
phase method}.

For this purpose it is enough to consider the ordinary integral:
\be
A(a,b)=\int^{+\infty}_{-\infty}\frac{dx}{(2\pi)^{1/2}}
e^{i(\frac{1}{2}ax^2+\frac{1}{3}bx^3)},
\l{3.14} \ee
with $Im~a \rar +0$ and $b>0$. Computing the `probability' $\R=|A|^2$
we find:
\ba
\R(a,b)=e^{\frac{1}{2i}\hat{j}\hat{e}}\int^{+\infty}_{-\infty} dx
e^{-2(x^2 +e^2 )Im~a}e^{2i\frac{b}{3}e^3}
\n\\\times\d (Re~ax +bx^2+j).
\l{3.1''}\ea
The `hat' symbol means the derivative over corresponding
quantity:  $\h{X}\equiv\pa/\pa X$. One should put the auxiliary
variables ($j,e$) equal to zero at the very end of calculations.

Performing the trivial transformation $e\rar ie$, $\hat{e}\rar
-i\hat{e}$ of auxiliary variable we find at the limit $\Im a=0$ that
the contribution of $x=0$ extremum (minimum) gives expression:
\be
\R(a,b)=\frac{1}{a}e^{-\frac{1}{2}\hat{j}\hat{e}}(1-4bj/a^2)^{-1/2}
e^{2\frac{b}{3}e^3}
\l{3.15'}\ee
and the expansion of operator exponent gives the asymptotic  series:
\be
\R(a,b)=\frac{1}{a}\sum^{\infty}_{n=0}(-1)^{n}\frac{(6n-1)!!}{n!}
(\frac{2b^4}{3a^6})^n.
\l{3.16'}\ee
This series is convergent in Borel sense.

Eq.(\r{3.1''}) can be considered as the definition of integral
(\r{3.14}).  By this reason one may put $\Im a =0$ from the very
beginning. We will use this property.

Let us calculate now $\R$ using the stationary phase method.
Contribution from the minimum $x=0$ gives $(\Im a=0)$:
$$
A(a,b)=e^{-i\hat{j}\hat{x}}e^{-\frac{i}{2a}j^2}e^{i\frac{b}{3}x^3}
(\frac{i}{a})^{1/2}.
$$
The corresponding `probability' is
\be
\R(a,b)=\frac{1}{a}e^{-\frac{1}{2}\hat{j}\hat{e}}e^{2\frac{b}{3}e^3}
e^{\frac{2b}{a^2}ej^2}
\l{3.17}\ee
This expression does not coincide with (\ref{3.15'}) but it leads to
the same asymptotic series (\r{3.16'}).

To find the representation (\r{3.17}) from (\r{3.15'}) the
transformation, see (\r{2.11'}),
\be
\d (\Re a~x +bx^2+j)=e^{-\f{i}{2}\h{j'}\h{e'}}e^{+2i(bx^2+j)e'}
\d (\Re a~x +j')
\l{trans}\ee
can be applied. Indeed, inserting this equality into (\r{3.15'}) we
find (\r{3.17}). The transformation (\r{trans}) becomes evident from
the Fourier transformation of $\d$-function. Eq.(\r{trans}) reflects
the freedom in choice of the perturbation theory in vicinity of
topologically-equivalent trajectories in functional space.\\

$\bullet$ The solution $x_j (t)$ of eq.(\r{3.13}) we would search
expanding it over $j(t)$:  $$ x_j (t)=x_c (t)+\int dt_1 G(t,t_1
)j(t_1 )+...  $$ This is sufficient since $j(t)$ is  the auxiliary
variable, i.e. we assume that $j$ is switched on adiabatically.  In
this decomposition $x_c (t)$ is the strict solution of unperturbated
equation $\ddot{x}+v'(x)=0$ and $G(t,t')$ must obey eq.(\r{2}). Note
that the functional $\d$-function in (\ref{*}) does not contain the
end-point values of time $t=0$ and $t=T$. This means that the initial
conditions to the eq.(\ref{3.13}) are arbitrary and the integration
over them is assumed because of our definition of $\R$.

The $\d$-likeness of measure allows to conclude:\\

{\it S8. All strict regular solutions (including trivial) of
classical (unperturbated by $j$) equation(s) of motion must be taken
into account.}

We must consider only `strict' solutions because of strict
cancellation of needless contributions when the $\d$-likeness of
measure is derived. The $\d$-likeness of measure means that the
probability $\R(E)$ should contain a $sum$ over all discussed
solutions.  This is the main distinction of our unitary method of
quantization from stationary phase method: even having few solutions
there is not interference terms in the sum over them in $\R$.

Note that the interference terms are absent independently from
solutions `nearness' in the functional space. This reflects the
orthogonality of Hilbert spaces builded on the various $x_c$ \C{gold}
and is the consequence of unitarity condition.

The solutions must be regular since the singular $x_c$ gives zero
contribution on $\d$-like measure.\\

$\bullet$ We offer following selection rule to define what
contribution is significant on the sum over various orbits $x_c$ :\\

{\it S9. In the sum over topologically nonequivalent trajectories one
should leave the contribution defined in the highest factor manifold
$G/\bar{G}$ if $G$ is the group violated by given $x_c$ and $\bar{G}$
is the $x_c$-invariance subgroup of $G$ group\footnote{Note, $G$
may be wider then the actions invariance group.}.}

Indeed, summation over all solutions of classical equation of motion
means necessity to take into account all topologically-equivalent
orbits $x_c$ also, i.e. means integration over parameters of factor
manifold $G/\bar{G}$.  This naturally introduces integration over
zero-mode degrees of freedom.  The corresponding measure will be
defined by mapping on the factor manifold $G/\bar{G}$, i.e. without
usage of the Faddeev-Popov {\it ansatz}.

It is evident that in the sum over contributions of various $x_c$ we
must leave largest, i.e. the term with maximal number of zero modes.
This selection rule \C{yad} presents our definition of the vacuum.\\

$\bullet$ The $\d$-like measure defines the real-time motion only and
is not applicable for tunnelling processes since reflects the
(space-)time-local equilibrium of all forces.  For instance, it
excludes the `kink' contributions for potential $v(x)\sim
(x^2-a^2)^2$.

This orbits belongs to the bifurcation line, using the terminology of
Smale \C{smale}, i.e.  the `kink' contributions should be added to
the contributions defined by our $\d$-like measure.  Then, following
to our selection rule, we should leave those contribution(s) which
are proportional to the highest zero-modes volume. So, our definition
of measure is rightful if the real-time contributions factor manifold
have the largest volume\footnote{One can say in this case that the
imaginary-time contributions are realized on zero measure.}.

The explicit investigation of this condition is the nontrivial task
in spite of its seeming simplicity (the volume of $G/\bar{G}$ is
defined by classical solution only). Actually we should know (i)
$all$ classical orbits and (ii) show that $G/\bar{G}$ is stable. So,
above selection rule gives the classification of mostly probable
contributions only.

It is evident that\\

{\it S10. The measure (\r{3.12}) admits the canonical
transformations}.

This evidently follows from $\d$-likeness of measure. The phase
space differential measure has the form:
\ba
DM(x,p)=\delta (E+\o -H_T (x))
\n\\\times
\prod_{t}dx dp
\delta(\dot{x}-\frac{\partial H_j}{\partial p})
\delta(\dot{p}+\frac{\partial H_j}{\partial x}),
\l{3.18}\ea
where
\be
H_{j}=\frac{1}{2}p^2 +v(x)-jx
\l{3.19}\ee
is the total Hamiltonian which is time dependent through $j(t)$.

We can introduce new pare $(\theta ,h)$ instead of $(x,p)$ inserting
\ba
1=\int D\th Dh\prod_{t}\d (h-\frac{1}{2}p^2 -v(x))\d (\th
\n\\
-\int^{x}dx(2(h-v(x)))^{-1/2}).
\l{3.20} \ea
It is important that both differential measures in (\r{3.20}) and
(\r{3.18}) are $\delta$-like. This allows to change the order of
integration surely and firstly integrate over $(x,p)$. Calculating
result one can use the $\d$-functions of (\r{3.18}). In this case
the $\d$-functions of (\r{3.20}) will define the constraints. But if
we will use the $\d$-functions of (\r{3.20}) the mapping $(x,p)\rar
(\th,h)$ is performed and the remaining $\d$-functions would define
motion in the factor space.  We conclude that our transformation
takes into account the constraints since both ways must give the same
result.

We find by explicit calculations that:
\be
DM(\th ,h)=\d (E+\o-h(T))\prod_{t}\d (\dot{\th}-\frac{\pa h_j}
{\pa h}) \d(\dot{h}+\f{\pa h_j}{\pa\th}),
\l{3.21}\ee
since considered transformation is canonical,
$\{h(x,p),\th (x,p)\}=1$, where
\be
h_j(\th,h) =h-jx_c (\th,h)
]\l{3.22}\ee
is the transformed Hamiltonian and $x_c (\theta,h)$ is the classical
trajectory parametrized by $h$ and $\th$.\\

$\bullet$ The $(\th,h)$ parametrized solution obey the equation:
$$\f{\pa x_c(\th,h)}{\pa \th}=p_c(\th,h),$$ see (\r{3.20}). One
should underline that only the non-trivial solution of this equation
is considered performing mapping $(x,p)\rar(\th,h)$. It is evident
that for trivial solution $x_c=0$, $p_c=0$ such transformation is
impossible since the corresponding cotangent manifold is empty.

The transformed perturbation theory presents expansion over $1/g$ if
$x_c\sim 1/g$, where $g$ is the interaction constant.  So, we wish
construct the perturbation theory in the `strong coupling' limit. But
one should remind also that all solutions must be taken into account.
This means that the perturbation theory for $\R(E)$ contains
simultaneously both series over $g$ (from trivial solution $x_c=0$)
and over $1/g$, i.e.  the sum of week-coupling and strong-coupling
expansions. According to our selection rule we should leave largest
among then.

On the cotangent bundle we must solve following equations of motion:
\be
\dot{h}=j\f{\pa x_c}{\pa\th},~~~\dot{\th}=1-j\f{\pa x_c}{\pa h}
\l{3.23} \ee
and they have a simple structure:\\

{\it S11. The Green function on the cotangent bundle is simple
$\Th$-function}.

Indeed, expanding solutions of eqs.(\r{3.23}) over $j$ in the zero
order we find $\th_0 =t_0 +t$ and $h_0 =const$. The first order gives
equation for Green function $g(t,t')$:  \be \dot{g}(t,t')=\d(t-t').
\l{3.24a}\ee The solution of this equation introduces the time
`irrevercibility':  \be g(t,t')=\Th (t-t'), \l{3.24}\ee in opposite
to causal particles propagator $G(t,t')$\footnote{Last one contains
sum of retarded and advanced parts.}. But, as will be seen below, see
$S10$, the perturbation theory with Green function (\r{3.24}) is time
reversible.  Note also, that the solution (\r{3.24}) is the unique
and is the direct consequence of usual in the quantum theories
$i\e$-prescription.

The uncertainty is contained in the boundary value $g(0)$. We will
see that $g(0)=0$ excludes some quantum corrections. By this reason
one should consider $g(0)\neq 0$. We will assume that \be g(0)=1
\l{3.24b}\ee since this boundary condition to eq.(\r{3.24a}) is
natural for local theories.  We will use also following formal
equalities:  \be g(t,t')g(t',t)=0,~~~1=g(t,t')+g(t',t) \l{3.24c}\ee
considering $g(t,t')$ as the distribution.\\

$\bullet$ It is important to note that $Im~g(t)=0$ on the real time
axis.  This allows to conclude that\\

{\it S12. The perturbation theory on the ($h,\th$) bundle can be
constructed on the real-time axis.}

Indeed, the $i\e$-prescription is not necessary since, as was
mentioned above in $S2$, the $\d$-functional measure defines a
complete set of contributions. But for more confidence one may
introduce the $i\e$-prescription and, extracting the $\d$-function in
the measure, one can put $\e =0$ if the contributions are regular at
this limit.

One can point out the examples when $\e =0$ is the singular point.

(a) The Green function G(t,t') is singular at $\e =0$. The
$i\e$-prescription introduces the wave damping in this case.

(b) The terms of perturbation theory are singular at $\e =0$ even if
the Green functions are regular. This singularities are connected
with light-cone singularities of the real-time theories.

(c) There is the tunneling phenomena. The $i\e$-prescription is
necessary to define a theory in the turning points (it is the usual
WKB prescription).\\

$\bullet$ Note now that $\partial x_c/\partial \theta$ and $\partial
x_c/\partial h$ in the r.h.s. of (\r{3.23})  can be  considered as
the sources. This allows to offer the statement:\\

{\it S13. The mapping on the cotangent bundle splits `Lagrange'
quantum force $j$ on a set of quantum forces individual to each
independent degree of freedom, i.e. to each independent local
coordinate of the cotangent manifold}.

Indeed, the simple algebra gives (see Appendix B):
\ba
R(E)=2\pi \int^{\infty}_{0}
dTe^{\frac{1}{2i}\hat{\omega}\hat{\tau}}
\n\\\times
e^{\frac{1}{2i}\Re\int_{C_+}dt
(\hat{j}_h (t)\hat{e}_h (t)+ \hat{j}_{\th} (t)\hat{e}_{\th} (t))}
\n\\ \times
\int Dh D\th e^{-i\tilde{H}(x_c ;\tau )-iV_T (x_c ,e_c )}
\delta (E+ \omega -h(T))
\n\\ \times
\prod_{t} \delta (\dot{h} -j_h )\delta
(\dot{\th} -1 - j_{\th}) \l{27b}\ea
Therefore, according to splitting $j \rightarrow (j_h ,j_{\th})$ we
must change $e \rightarrow e_c$, where \be e_c =e_h \frac{\pa
x_c}{\pa \th} -e_{\th} \frac{\pa x_c}{\pa h} \equiv (e_h
\h{\th}-e_\th\h{h})x_c.  \l{23b}\ee carry the symplectic structure of
Hamilton's equations of motion, see (\r{3.21}), i.e. $e_c$ is the
invariant of canonical transformations. This quantity describes the
flow $\d_{h}x_c \bigwedge \d_{\th}p_c$ generated by quantum
perturbations through the bundles elementary cell.

Hiding the $x_c (t)$ dependence in $e_c$ we had solve the problem of
the functional determinants and simplify the equation of motion as
much as possible:  \ba DM(h,\th )=\d (E+\o -h(T)-h'(T))
\n\\\times\prod_{t}
dh(t) d\th(t) \d (\dot{h}(t)) \d (\dot{\theta}(t)-1) \l{3.26}\ea and
the perturbations generating operator \ba
\h{K}=\frac{1}{2}(\hat{\o}\hat{\tau}+ \int^T_0 dt_1 dt_2 \Theta (t_1
-t_2 ) (\hat{e}_h (t_1 )\hat{h}'(t_2 )\n\\+ \hat{e}_{\theta}(t_1
)\hat{\theta}' (t_2 )).  \l{3.27}\ea In $U_T (x_c,e_c)$ we must
change $h \rightarrow (h + h')$ and $\th \rightarrow (\th +\th')$.

Noting that $$ \int \prod_t dX(t)\d (\dot{X}(t))=\int dX(0)=\int dX_0
$$ we see that the measure (\r{3.26}) coincide with the measure of
ordinary integrals over $h_0$ and $t_0$. Last one defines the
volume of translational mode.\\

$\bullet$ Let us consider motion in the action-angle phase space.
Corresponding perturbations generating operator has the form:  \ba
\h{K}=\frac{1}{2}\int_0^T dtdt' \Theta (t'-t) (\hat{I}(t)\hat{e}_I
(t')+\hat{\phi}(t)\hat{e}_{\phi} (t'))\n\\\equiv \h{K}_I +\h{K}_{\p}.
\l{3.32}\ea The result of integration using last $\d$-function is \be
R(E)=2\pi \int^{\infty}_{0} dT  e^{-i\h{K}}\int^{2\pi}_{0}
\frac{d\phi_0}{\o (E)}e^{-iU_T (x_c ,e_c )}, \l{3.33}\ee where
$$\o =\pa h(I_0)/\pa I_0$$ and $I_0=I_0(E)$ is defined by the
algebraic equation:  $$E=h(I).$$ The classical trajectory \be x_c
(t)=x_c (I_0(E)+I(t)-I(T), \phi_0 +\tilde{\o}t+\phi (t)),
\l{3.33'}\ee where $$\tilde{\o}=\f{1}{t}\int dt' g(t,t')\o (I_0
+I(t')).$$ The interaction `potential' $V_T$ depends from \be
e_c=e_{\p}\f{\pa x_c}{\pa I}- e_{I}\f{\pa x_c}{\pa \p}.
\l{3.33''}\ee

One can note that eq.(\ref{3.29}) contains unnecessary contributions.
Indeed, action of the operator $$ \int^{T}_{0}dt dt' \Theta (t-t')
\hat{e}_I (t)\hat{I}(t') $$ on $\tilde{H}(x_c ;\tau )$, defined in
(\ref{3.7'}), leads to the time integrals with zero integration range:
$$ \int^{T}_{0}dt \Theta (T-t) \Theta (t-T) =0.  $$ This
simplification was used in (\r{3.32}) and (\r{3.33}).

The operator $\h{K}$ is linear over $\h e_{\p}$, $\h e_{I}$. The
result of its action can be written in the form:  \be R(E)=2\pi
\int^{\infty}_{0} dT \int^{2\pi}_{0} \frac{d\phi_0}{\o
(E)}:e^{-iU_T (x_c ,\h{e}_c/2i)}:, \l{a1}\ee where \ba
\h{e}_c=\h{j}_{\p}\f{\pa x_c}{\pa I}- \h{j}_{I}\f{\pa x_c}{\pa \p}=
(\h{j}_\p\h{I}-\h{j}_I\h{\p})x_c\n\\=\int^T_0 dt' \th(t-t')
\{\h{\p}(t'),\h{I}(t)\}x_c(t) \l{a2}\ea since \be \h{j}_{X}(t)=\int^T
_0 dt' \th (t-t') \h{X}(t'),~~~X=\p ,I.  \l{a3}\ee The colons in
(\r{a1}) means 'normal product': the differential operators must stay
to the left of $all$ functions in expansion over commutator
$$\{\h{\p}(t'),\h{I}(t)\}=\h{\p}(t')\h{I}(t)-\h{I}(t')\h{\p}(t).$$

Now we are ready to offer the important statement:\\

{\it S14. If the eqs.(\r{3.24}, \r{3.24b}, \r{3.24c}) are hold then
each term of perturbation theory in the invariant subspace
can be represented as the sum of total derivatives over the subspace
coordinates.}

This statement directly follows from definition of perturbation
generating operator $\h{K}$ on the cotangent bundle (\r{3.27}) and of
translationally invariance of the cotangent manifold in the classical
approximation. The proof of this statement is given in Appendix C.

We can conclude, contributions are defined by boundary values of
classical trajectory $x_c$ in the invariant subspace since the
integration over $X_0$ is assumed, see (\r{3.33}), and since
contributions are the total derivatives over $X_0$. \\

$\bullet$ One can observe following new phenomena:\\

{\it S15. The quantum corrections to angular variables are canceled
if the classical motion is periodic.}

If $x_c$ is the periodic function:
\ba
x_c (I_0(E)+I(t)-I(T),
(\phi_0 +2\pi ) +\tilde{\Omega}t+\phi (t))= \n\\x_c
(I_0(E)+I(t)-I(T), \phi_0 +\tilde{\Omega}t+\phi (t)).
\l{3.37} \ea
this statement is elementary consequence of $S10$ and is the result
of averaging over $\p_0$, see eq.(\r{3.33}).

We would like to note now that, generally speaking,\\

{\it S16. The transformed measure can not be deduced from direct
transformations of path integral (\r{3.1})}.

Let us consider now the coordinate transformations. For instance, the
two dimensional model with potential
$v=v((x^{2}_{1}+x^{2}_{2})^{1/2})$ is simplified considering it in
the cylindrical coordinates $x_1 =r \cos\phi$, $x_2 =r \sin\phi$.
Note, this transformation is not canonical.

Starting from flat space with trivial metric tensor $g_{\mu\nu}$ and
inserting
\be
1=\int Dr D\p \prod_t \d
(r-\sqrt{x_1^2 +x_2^2}) \d (\p - \arctan \f{x_2}{x_1})
\l{coord}\ee
we find the measure in the cylindrical coordinates:
\ba
D^{(2)}M(r,\p )=\d (E+\o -H_T (r,\p))\times
\n\\
\prod_t dr d\p r^2 (t) \d (\ddot{r}-\dot{\p}^2 r+v'(r)-j_r)\d (\pa_t
(\dot{\p}r^2 )-rj_{\p}),
\l{3.28}\ea
where $v'(r)=\pa v(r)/\pa r$ and $j_r$, $j_{\p}$ are the components
of $\vec{j}$ in the cylindrical coordinates.

The perturbation generating operator has the form:
\be
\h{K}(j,e)=\frac{1}{2}\{\hat{\o}\hat{\tau}+Re\int_{C_+} dt
(\hat{j}_{r}(t)\hat{e}_{r}(t)+\hat{j}_{\phi}(t)\hat{e}_{\phi}(t))\}
\l{3.29}\ee
and in $U_T (x,\vec{e})$ we must change $\vec{e}$ on $\vec{e}_c$
with components
\be
e_{c,1} =e_r \cos\p -re_{\p}\sin\p,~~~
e_{c,2}=e_r \sin\p +re_{\p} \cos\p.
\l{3.30}\ee
Note, $e_\p$ was arise in product with $r$. To find (\r{3.29}) and
(\r{3.30}) one can use (\r{26'})

The transformation looks quite classically but the measure (\r{3.28})
and perturbation generating operator (\r{3.29}) can not be derived by
$naive$ coordinate transformation of initial path integral for
amplitude. This becomes evident noting that transformed
representation for $\R_1(E)$ can not be written in the product form
$\sim AA^*$ of two functional integrals.

Indeed, $\R_1(E)$ has not the factorization property because of the
mixing of various quantum degrees of freedom when the transformation
was performed. This is seen explicitly calculating $\R_1(E)$ with the
measure (\r{3.28}) and the perturbations generating operator
(\r{3.29}):
\ba
\R_1(E)=4\pi\int^{\infty}_0 dT d\tau e^{2iE\tau}
\n\\\times
\int\prod_t\f{r^2drd\p}{\pi^2}de_rde_\p
e^{iS_{T+\tau}(x+e)-iS_{T-\tau}(x-e)}
\n \\\times
e^{-2i\Re\int dt[\f{\d S_T (x)}{\d x}\cdot e
-\f{\d S_T (r,\p)}{\d r}e_r -
\f{\d S_T (r,\p)}{\d \p}e_p ]},
\l{i1}\ea
where the action of $\exp\{-i\h K\}$ was performed. It is assumed that
all quantities in this expression are written in the cylindrical
coordinates.

Introducing the `main' variables $r_\pm =r \pm e_r$, $\p_\pm =\p
\pm e_\p$ and $T_\pm =T \pm \tau$ we can see easily that (\r{i1}) can
not be factorised onto product of two path integrals. For instance,
\ba
\prod_t\f{r^2drd\p}{\pi^2}de_rde_\p=\prod_t (r_+ +r_-)^2
\f{dr_+d\p_+}{2\pi}\f{dr_-d\p_-}{2\pi}\neq
\n\\
\{\prod_t r_+\f{dr_+d\p_+}{2\pi}\}\{\prod_t r_-\f{dr_-d\p_-}{2\pi}\}
\ea

$Note$, we can introduce also the motion in the phase space with
Hamiltonian $$H_{j}=\f{1}{2}p^2
+\f{l^2}{2r^2}+v(r)-j_{r}r-j_{\p}\p.$$ The Dirac measure becomes four
dimensional:  \ba D^{(4)}M(r,\p,p,l) =\d(E+\o -H_T)
\n\\\times
\prod_tdr(t)d\p(t)dp(t)dl(t)
\n\\\times
\d(\dot{r}-\frac{\pa H_j}{\pa p})\d(\dot{\p}-\f{\pa H_j}{\pa l})
\d(\dot{p}+\f{\pa H_j}{\pa r})\d(\dot{l}+\f{\pa H_j}{\pa\p}).
\l{3.31}\ea Note absence of the coefficient $\prod r^2(t)$ in this
expression.

It is interesting also to find the measure starting from the curved
space with the Lagrangian
\be
L=\f{1}{2}g_{\mu\nu}(y)\dot{y}^\mu\dot{y}^\nu -{v}(y)
\l{b6'}\ee
It is enough to consider the kinetic term only since, to find the
Dirac measure, we should extract the odd over $e$ terms from the
`closed-path' action $S_T (y+e)-S_T (y-e)$. This procedure is
`trivial' for potential term. The lowest over $e^\la$ part of the
kinetic term have the form:
\be
2\{g_{\la\mu}\ddot{y}^\mu +\Ga_{\la,\mu\nu}\dot{y}^\mu\dot{y}^\nu\}.
\l{b7}\ee
Therefore, the quasiclassical approximation is restored.

To find the quantum corrections we should linearise at least the
$O(e^3)$ term in the exponent
$$
\exp\{\Re\int dt g_{\la,\mu\nu} e^\la
\dot{e}^\mu\dot{e}^\nu\}.
$$
using (\r{26'}). This is possible noting that
$$
e^\mu(t')\h{e}'_\mu(t')\dot{e}'^\nu(t)=e^\mu(t')\d_{\mu\nu}
\pa_{t'}\d(t-t')=\dot{e}^\nu \d(t-t').
$$
In result,
\ba
DM(y)=\sqrt{|g(y+e)||g(y-e)|}\prod_\la\prod_t dy_\la \d
(g_{\la\mu}\ddot{y}^\mu
\n\\
+\Ga_{\la,\mu\nu}\dot{y}^\mu\dot{y}^\nu
+v_\la (y) -j_\la).
\l{b8}\ea
where $v_\la (y)=\pa_\la v(y)$ and $\Ga_{\la,\mu\nu}$ is the
Christoffel index.  The perturbations generating operator $\h{K}$ and
the weight functional $V(y;e)$ have the standard form.\\

$\bullet$ $Note$, above consideration shows that the mechanical
systems quantization in the space of nontrivial topology crucially
depends from the way as the metrics is introduced.

\section{H-atom problem}\setcounter{equation}{0}

We will calculate the integral:
\be
\R_1(E)=\int^{\infty}_0 dTe^{-i\h{K}(j,e)}\int DM(p,l,r,\vp)
e^{-iU_T(r,e)},
\l{21}\ee
where $\R_1(E)$ is the $probability$ to find a particle with energy
$E$, i.e.  we should find \C{yad} that normalized on the zero-modes
volume
\be
\R_1(E)=\pi\sum_n \d (E-E_n),
\l{22}\ee
where $E_n$ are the bound states energies. For $H$-atom problem
$E_n\leq 0$.  This condition would define considered homotopy class.

Expansion over operator
\be
\h{K}(j,e)=\f{1}{2}\int^T_0 dt (\h{j}_r\h{e}_r +
\h{j}_{\vp}\h{e}_{\vp}),~~~
\h{X}(t)\equiv \d/\d X(t),
\l{23}\ee
generates the perturbation series. It will be seen that in our case
we may omit the question of perturbation theories convergence.

The differential measure
\ba
DM(p,l,r,\vp)=\d (E-H_0)
\n\\\times
\prod_t dr(t) dp(t) dl(t) d\vp (t)
\n \\ \times
\d (\dot{r}-\f{\pa H_j}{\pa p})
\d (\dot{p}+\f{\pa H_j}{\pa r})
\d (\dot{\vp}-\f{\pa H_j}{\pa l})
\d (\dot{l}+\f{\pa H_j}{\pa \vp}),
\l{24}\ea
with total Hamiltonian ($H_0=H_j|_{j=0}$)
\be
H_j=\f{1}{2}p^2 -\f{l^2}{2r^2}-\f{1}{r}-j_r r -j_{\vp}\vp
\l{25}\ee
allows perform arbitrary transformations because of its
$\d$-likeness. The functional
\ba
U_T(r,e)=-s_0(r)+
\n\\
+\int^T_0 dt[\f{1}{((r+e_r)^2+r^2e_{\vp}^2)^{1/2}}
\n\\
-\f{1}{((r-e_r)^2+r^2e_{\vp}^2)^{1/2}}+2\f{e_r}{r}]
\l{26}\ea
describes the interaction between various quantum modes and $s_0 (r)$
defines the nonintegrable phase factor \C{yad}. The quantization of
this factor determines the bound state energy (see below).  Such
factor will appear if the phase of amplitude can not be fixed (as,
for instance, in the Aharonov-Bohm case).  Note that the Hamiltonian
(\r{25}) contains the energy of radial $j_r r$ and angular
$j_{\vp}\vp$ excitation independently.

We would like to offer following general method of mapping. It is
important start from the  assumption that the invariant subspace has
symplectic structure of cotangent manifold $T^*G$ and its farther
possible reduction to linear subspace $W$ ($dim(T^*G)\geq dim(W)$)
would be realized as the reduction of quantum degrees of
freedom\footnote{This reminds the classic reduction phenomena since
the `quantum reduction' arise if additional degrees of freedom of
$T^*G$ are the irrelevant variables for classical flow.}.

Therefore, the first step of mapping consist in demonstration that \\

{\it S17. The classical trajectories belong to $T^*G$ completely.}

Let
\ba
\D=\int \prod_t d^2\x d^2\eta \d (r-r_c(\x,\eta))\d (p-p_c(\x,\eta))
\n\\\times
\d (l-l_c(\x,\eta))\d (\vp-\vp_c(\x,\eta))
\l{27}\ea
be the functional of known functions $(r_c,p_c,\vp_c,l_c)(\x,\eta)$,
where $(\x,~\eta)$ are two-vectors. It is assumed that one can find
such functions $(\x,\eta)(t)$ at given $(r,p,\vp,l)(t)$ that the
functional determinant
\ba
\D_c=\int \prod_t d^2\bar{\x} d^2\bar{\eta}
\d(\f{\pa r_c}{\pa\x}\cdot\bar{\x}+\f{\pa r_c}{\pa\eta}
\cdot\bar{\eta})
\n \\\times
\d(\f{\pa p_c}{\pa\x}\cdot\bar{\x}+
\f{\pa p_c}{\pa\eta}\cdot\bar{\eta})
\d(\f{\pa \vp_c}{\pa\x}\cdot\bar{\x}+\f{\pa \vp_c}{\pa\eta}
\cdot\bar{\eta})
\n\\\times
\d(\f{\pa l_c}{\pa\x}\cdot\bar{\x}+
\f{\pa l_c}{\pa\eta}\cdot\bar{\eta})\neq 0.
\l{28}\ea
Note that this is the condition for $(r_c,p_c,\vp_c,l_c)(\x,\eta)$
only since one can choose $(r,p,\vp,l)(t)$ in eq.(\r{27}) in an
arbitrary useful way.

To perform the mapping we should insert
$$
1=\D/\D_c
$$
into (\r{21}) and integrate over $r(t)$, $p(t)$, $\vp(t)$ and $l(t)$.
In result of simple calculations (see Appendix D) we find that
\be
DM(\x, \eta)=\d(E-H_0)\prod_td^2\x d^2\eta
\d^2(\dot{\x}-\f{\pa h_j}{\pa\eta})
\d^2(\dot{\eta}+\f{\pa h_j}{\pa \x}),
\l{212'}\ee
It is the desired result of transformation of the measure for given
`generating' functions $(r_c,p_c,\vp_c,l_c)(\x,\eta)$. In this case
the `Hamiltonian' $h_j (\x,\eta)$ is defined by four equations
(\r{211}).  But there is another possibility. Let us assume that
\be
h_j (\x, \eta)=H_j (r_c, p_c, \vp_c, l_c)
\l{213}\ee
and the functions $(r_c,p_c,\vp_c,l_c)(\x,\eta)$ are unknown. Then
eqs.(\r{211}) are the equations for this functions. It is not hard to
see that the eqs.(\r{211}) simultaneously with equations given by
$\d$-functions in (\r{212'}) are equivalent of incident equations if
the equality (\r{213}) is hold. So, incident dynamical problem was
divided on two parts. First one defines the trajectory in the $W$
space through eqs.(\r{211}). Second one defines the dynamics, i.e.
the time dependence, through the equations in arguments of
$\d$-functions in the measure.\\

$\bullet$ Therefore, we should consider $r_c,~ p_c,~ \vp_c,~ l_c$ as
the solutions in the $\x,~\eta$ parametrization. The desired
parametrization of classical orbits has the form (one can find it in
arbitrary textbook of classical mechanics):
\ba
r_c=\f{\eta_1^2(\eta_1^2+\eta_2^2)^{1/2}}
{(\eta_1^2+\eta_2^2)^{1/2}+\eta_2\cos \x_1},~\vp_c=\x_1,
\n\\
p_c=\f{\eta_2\sin \x_1}{\eta_1(\eta_1^2+\eta_2^2)^{1/2}},~l_c=\eta_1.
\l{214}\ea
At the same time,
\be
h_j=\f{1}{2(\eta_1^2+\eta_2^2)^{1/2}} -j_r r_c -j_\vp \x_1
\equiv h (\eta)-j_r r_c -j_\vp \x_1.
\l{215}\ee
Note that $\x_2$ is the irrelevant variable for classical flow
(\r{214}). This conclusion hides the assumption that the space is
flat and homogeneous. So, the external field would violate our
solution.

Noting that the derivatives over $\x_2$ are equal to
zero\footnote{To have the condition (\r{28}) we should assume that
$\pa r_c/\pa \x_2 \sim \e \neq 0$. We put $\e =0$ at the end of
transformation.} we find that
\ba
DM(\x, \eta)=\d(E-h(T))\prod_td^2\x
d^2\eta \d(\dot{\x}_1-\o_1+j_r\f{r_c}{\pa\eta_1}) \n \\ \times
\d(\dot{\x}_2-\o_2+j_r\f{r_c}{\pa\eta_2})
\d(\dot{\eta}_1-j_r\f{\pa r_c}{\pa \x_1} -j_\vp)
\d(\dot{\eta}_2),
\l{217}\ea
where
\be
\o_i=\pa h/\pa\eta_i
\l{218}\ee
are the conserved in classical limit $j_r=j_\vp =0$ `velocities' in
the $W$ space.

Now we can show the example of reduction of quantum degrees of
freedom. We can conclude that\\

{\it S18. The dynamical variable stay the $c$-number if it has not
the canonically conjugate pare.}

We see from (\r{217}) that the length of Runge-Lentz vector is not
perturbated by the quantum forces $j_r$ and $j_{\vp}$. To investigate
the consequence of this fact it is useful to project this forces on
the axis of $W$ space. This means splitting of $j_r,~j_{\vp}$ on
$j_\x,~j_\eta$.  The equality
\ba
\prod_t\d(\dot{\x}_1-\o_1+j_r\f{r_c}{\pa\eta_1})=
e^{\f{1}{2i}\int^T_0 dt \h{j}_{\x_1}\h{e}_{\x_1}}
\n\\\times
e^{2i\int^T_0 dt j_r e_{\x_1}\pa r_c/\pa \eta_1}
\prod_t\d(\dot{\x}_1-\o_1+j_{\x_1})
\ea
becomes evident if the Fourier representation of $\d$-function is
used, $S13$ (see also \C{yaph}). The same transformation of arguments
of other $\d$-functions in (\r{217}) can be applied. Then, noting
that the last $\d$-function in (\r{217}) is source-free, we find the
same representation as (\r{21}) but with \be \h{K}(j,e)=\int^T_0 dt
(\h{j}_{\x_1}\h{e}_{\x_1}+ \h{j}_{\x_2}\h{e}_{\x_2}+
\h{j}_{\eta_1}\h{e}_{\eta_1}),
\l{31}\ee
where the operators $\h{j}$ are defined by the equality:
\be
\h{j}_X (t)=\int^T_0 dt' \theta(t- t')\h{X}(t')
\l{44}\ee
and $\theta(t- t')$ is the Green function of our perturbation theory
\C{yaph}.

We should change also
\be
e_r\rar e_c=e_{\eta_1}\f{\pa r_c}{\pa \x_1}-
e_{\x_1}\f{\pa r_c}{\pa \eta_1}-
e_{\x_2}\f{\pa r_c}{\pa \eta_2},~~e_\vp\rar e_{\x_1}
\l{32}\ee
in the eq.(\r{26}). The differential measure takes the simplest form:
\ba
DM(\x, \eta)=\d(E-h(T))
\n\\\times
\prod_td^2\x d^2\eta
\d(\dot{\x}_1-\o_1-j_{\x_1})
\d(\dot{\x}_2-\o_2-j_{\x_2})
\n \\\times
\d(\dot{\eta}_1-j_{\eta_1})
\d(\dot{\eta}_2).
\l{33}\ea

Note now that the $\x, \eta$ variables are contained in $r_c$ only:
$$
r_c= r_c (\x_1, \eta_1, \eta_2).
$$
This means that the action of the operator $\h{j}_{\x_2}$ gives
identical to zero contributions into perturbation theory series. And,
since $\h{e}_{\x_2}$ and $\h{j}_{\x_2}$ are conjugate operators, see
(\r{31}), we can put
$$
j_{\x_2}=e_{\x_2}=0.
$$
This conclusion ends the reduction:
\be
\h{K}(j,e)=\int^T_0 dt (\h{j}_{\x_1}\h{e}_{\x_1}+
\h{j}_{\eta_1}\h{e}_{\eta_1}),
\l{34}\ee
\be
e_c=e_{\eta_1}\f{\pa r_c}{\pa \x_1}-e_{\x_1}\f{\pa r_c}{\pa \eta_1}.
\l{34'}\ee
The measure has the form:
\ba
DM(\x, \eta)=\d(E-h(T))d\x_2 d\eta_2
\n\\\times
\prod_t d\x_1 d\eta_1
\d(\dot{\x}_1-\o_1-j_{\x_1})
\d(\dot{\eta}_1-j_{\eta_1})
\l{35}\ea
since $U_T=U_T(r_c, e_c, \x_1)$ is $\x_2$ independent and
$$
\int \prod_t dX(t)\d(\dot{X})=\int dX(0).
$$

$\bullet$ One can see from (\r{35}) that the reduction can not solve
the H-atom problem completely: there are nontrivial corrections to
the orbital degrees of freedom $\x_1,\eta_1$. By this reason we
should consider the expansion over $\h{K}$. In result we will see
that\\

{\it S19. The quantum corrections may give zero contributions if the
interactions disappeared on the bifurcation line.}

Using last $\d$-functions in (\r{35}) we find, see also \C{yaph}
(normalizing $\R_1(E)$ on the integral over $\x_2$):
\be
\R(E)=\int^\infty_0 dT e^{-i\h{K}(j,e)}\int dM e^{-iU_T(r_c,e)},
\l{41}\ee
where
\be
dM=\f{d\x_1 d\eta_1}{\o_2(E)}.
\l{42}\ee
The operator $\h{K}(j,e)$ was defined in (\r{34}) and
\ba
U_T(r,e)=-s_0(r)
+\int^T_0 dt[\f{1}{((r_c+e_c)^2+
r_c^2e_{\x_1}^2)^{1/2}}
\n\\
-\f{1}{((r_c-e_c)^2+r_c^2e_{\x_1}^2)^{1/2}}
+2\f{e_c}{r_c}]
\l{a}\ea
with $e_c$ defined in (\r{34'}) and
\ba
r_c(t)=r_c(\eta_1 +\eta(t), \bar{\eta}_2(E,T), \x_1+\o_1(t)
+\x(t)),
\n\\
E\equiv h(\eta_1 +\eta(T), \bar{\eta}_2),
\l{43}\ea
where $\bar{\eta}_2(E,T)$ is the solution of equation $E=h$.

The integration range over $\x_1$ and $\eta_1$ is as follows:
\be
\pa W_C: 0\leq \x_1 \leq 2\pi,~~-\infty \leq \eta_1 \leq +\infty.
\l{45}\ee
First inequality defines the principal domain of the angular variable
$\vp$ and second ones take into account the clockwise and
anticlockwise motions of particle on the Kepler orbits, $|\eta_1|=
\infty$ is the bifurcation line.

We can write:
\be
\R(E)=\int^\infty_0 dT \int dM :e^{-iV(r_c,\h{e})}:
\l{47}\ee
since the operator $\h{K}$ is linear over $\h{e}_{\x_1},
\h{e}_{\eta_1}$.  The colons means `normal product' with operators
staying to the left of functions and $V(r_c,\h{e})$ is the functional
of operators:
\be
2i\h{e}_c=\h{j}_{\eta_1}\f{\pa r_c}{\pa \x_1}-
\h{j}_{\x_1}\f{\pa r_c}{\pa \eta_1},~~2i\h{e}_{\x_1}=\h{j}_{\x_1}.
\l{48}\ee
Expanding $U_T(r_c, \h{e})$ over $\h{e}_c$ and $\h{e}_{\eta_1}$ we
find:
\be
U_T(r_c,\h{e})=-s_0(r_c) +2\sum_{n+m \geq 1}C_{n,m}\int^T_0 dt
\f{\h{e}_c^{2n+1}\h{e}_{\eta_1}^m}{r_c^{2n+2}},
\l{46}\ee
where $C_{n,m}$ are the numerical coefficients. We see that the
interaction part presents expansion over $1/r_c$ and, therefore, the
expansion over $V$ generates an expansion over $1/r_c$.

In result,
\ba
\R(E)=\int^\infty_0 dT \int dM \{e^{is_0 (r_c)} +
B_{\x_1}(\x_1, \eta_1)
\n\\
+B_{\eta_1}(\x_1, \eta_1)\}.
\l{49}\ea
The first term is the pure quasiclassical contribution and last ones
are the quantum corrections. They can be written as the total
derivatives:
\be
B_{\x_1}=\f{\pa}{\pa
\x_1}b_{\x_1},~ B_{\eta_1}=\f{\pa}{\pa\eta_1}b_{\eta_1}.
\l{410}\ee
This means that the mean value of quantum corrections in the $\x_1$
direction are equal to zero:
\be
\int^{2\pi}_0 d\x_1 \f{\pa}{\pa \x_1}b_{\x_1}(\x_1,
\eta_1) =0
\l{411}\ee
since $r_c$ is the closed trajectory independently from initial
conditions, see $S15$.

In the $\eta_1$ direction the motion is classical:
\be
\int^{+\infty}_{-\infty} d\eta_1 \f{\pa}{\pa \eta_1}
b_{\eta_1}(\x_1, \eta_1)=0
\l{412}\ee
since (i) $b_{\eta_1}$ is the series over $1/r_c^2$ and (ii) $r_c
\rar \infty$ when $|\eta_1| \rar \infty$. Therefore,
\be
\R(E)=\int^\infty_0 dT \int dM e^{is_0 (r_c)}.
\l{413}\ee
This is the desired result.\\

$\bullet$ Noting that
$$
s_0 (r_c)= kS_1 (E),~~k=\pm 1, \pm 2,...,
$$
where $S_1 (E)$ is the action over one classical period $T_1$:
$$
\frac{\partial S_1 (E)}{\partial E}=T_1 (E),
$$
and using the identity \C{yad}:
$$
\sum^{+\infty}_{-\infty} e^{inS_1 (E)} =
2\pi \sum^{+\infty}_{-\infty}\d (S_1 (E) - 2\pi n),
$$
we find normalizing on zero-modes volume, that
\be
\R(E)=\pi \sum_{n} \d (E + 1/2n^2).
\l{416}\ee

\section{sin-Gordon model}\0

Our aim is to calculate the integral:
\be
\R_2 (q) = e^{-i \hat K (j,e)}\int DM (u,p)
|\Ga (q;u)|^2 e^{is_0 (u)-i U(u,e)},
\l{2.1}\ee
where $\Ga (q;u)$ was defined in (\r{1.2}). The time integrals would
be defined on the Mills time contour \C{mil} to avoid the mass-shell
singularities of the perturbation theory.

In this expression the expansion over operator
\ba
\hat K (j,e)=\f{1}{2}\Re\int_{C_+} dx dt \f{\d}{\d j(x,t)}\f{\d}{\d
e(x,t)}
\n\\
\equiv \f{1}{2}\Re\int_{C_+} dx dt \hat{j}(x,t)\hat{e}(x,t)
\l{2.2'}\ea
generates the perturbation series. We will assume that
this series exist.
The variational derivatives in (\r{2.2'}) are defined as follows:
$$
\f{\d \phi(x,t\in C_i)}{\d \phi(x',t'\in C_j)}=\d_{ij}\d
(x-x') \d (t-t'),~~~ i,j=+,-.
$$ The auxiliary variables $(j,e)$ must be taken equal to zero at the
very end of calculations.

The functionals $U(u,e)$ and $s_0(u)$ are defined
by the equalities:
\ba
U(u,e)= (V(u+e) - V(u-e))
\n\\
-2\Re\int_{C_+} dx dte(x,t) v'(u) ,
\n \\
s_0(u)=(S_0(u+e)-S_0(u-e))
\n\\
+2\Re\int_{C_+}dx dte(x,t)(\pa^2+m^2)u(x,t),
\l{2.3}\ea
where $S_0(u)$ corresponds to the free part of Lagrangian (\r{1.13})
and $V(u)$ describes interactions. The quantity $s_0 (u)$ is not
equal to zero since soliton configurations have nontrivial
topological charge (see also \C{yad}).

Considering motion in the phase space $(u,p)$ the measure
$DM(u,p)$ has the form:
\ba
DM(u,p)=
\prod_{x,t}
du(x,t) dp(x,t)
\n\\\times
\d (\dot{u}-\frac{\d H_j (u,p)}{\d p})
\d (\dot{p}+\frac{\d H_j (u,p)}{\d u})
\l{2.4}\ea
with the total Hamiltonian
\be
H_j(u,p)=\int dx \{ \frac{1}{2} p^2 +\frac{1}{2} (\pa_x u)^2
- \f{m^2}{\la^2}[\cos(\la u)-1] -ju \}.
\l{2.5}\ee

The problem will be considered assuming that $u(x,t)$ belongs to
Schwarz space:
\be
u(x,t)|_{|x| =\infty}=0~(mod \frac{2\pi}{\la}).
\l{2.7}\ee
This means that $u(x,t)$ tends to zero $(mod \frac{2\pi}{\la})$ at
$|x|\rar \infty$ faster then any power of $1/|x|$.

The measure (\r{2.4}) allows to perform arbitrary transformations.
But, as was explained in Sec., the canonical transformations are
essential since they conserve the form of equations of motion and
help to define the quantum degrees of freedom in the $W$ space, see
$S $.

Hence, assuming that this transformation exist \C{takh}, one may
propose that\\

{\it S21. The $N$-solitons functional measure in the $W$ space has
the form:}
\ba
D^N M (\x ,\eta)=\prod_{t}d^N \x (t) d^N
\eta (t)
\n\\\times
\d (\dot{\x}-\frac{\pa h_j (\x ,\eta )}{\pa \eta (t)}) \d
(\dot{\eta}+\frac{\pa h_j (\x ,\eta )}{\pa \x (t)}),
\l{2.8}\ea
where $h_j$ is the transformed Hamiltonian:
\be
h_j(\x ,\eta ) =h_N (\eta) -\int dx j(x,t) u_N (x;\x ,\eta )
\l{2.9}\ee
and $u_N (x;\x ,\eta )$ is the $N$-soliton configuration parametrized
by $(\x ,\eta )$.

The proof of eq.(\r{2.8}) is the same as for considered above
Coulomb problem. But the case of $(1+1)$-dimensional space needs
additional explanations. First of all one must introduce
\ba
\D(u,p)=\int \prod_t d^N \x (t) d^N \eta (t)
\n\\\times
\prod_{x,t}\d (u(x,t)-u_c(x;\x ,\eta))
\d (p(x,t)-p_c(x;\x ,\eta))
\l{2.22}\ea
This distribution is infinite. But this infinity is formal because of
disappearance of the determinant of transformation $\D(u_c,p_c)$ in the
final result iff the Poisson brackets:
\be
\{u_c(x,t),h_j\}=\f{\d H_j}{\d p_c(x,t)},~~~
\{p_c(x,t),h_j\}=-\f{\d H_j}{\d u_c(x,t)}.
\l{2.17'}\ee
or, using definition (\r{2.9}), iff
\ba
\{u_c(x;\x ,\eta),u_c(y;\x ,\eta)\}
\n\\= \{p_c(x;\x
,\eta),p_c(y;\x ,\eta)\}=0,
\n \\
\{u_c(x;\x ,\eta),p_c(y;\x ,\eta)\}=\d (x-y)
\l{2.23}\ea
are hold. For more confidence one can introduce the cells in the
$x$-space \C{takh}. Eqs.  (\r{2.23}) are the necessary and
sufficient conditions for considered mapping

\be
J:~\{u,p\}(x,t)\rar\{\x,\eta\}(t)
\l{sGm}\ee
with local coordinates $(\x ,\eta)$ defined by the equations:
\be
\dot{\x} =\f{\pa h_j}{\pa \eta},~~~
\dot{\eta} =-\f{\pa h_j}{\pa \x}.
\l{2.24'}\ee
Eqs.(\r{2.23},~\r{2.24'}) must be considered simultaneously.\\

$\bullet$ The eqs.(\r{2.17'}) fulfilled for arbitrary $j(x,t)$.
Therefore, the quantum perturbations should not alter the Poisson
brackets algebra.  In our terms this means that the quantum force
$j(x,t)$ excites the $(\x, \eta)$ manifold only, leaving the topology
of classical trajectory $(u,p)_c$ unchanged. So, since the complete
set of canonical coordinates $(\x ,\eta)$ for sin-Gordon model is
known, see e.g.  \C{takh}, we can use them immediately.

The classical Hamiltonian $h$ is the sum:
\be
h(\eta)=\int dp \s (r)\sqrt{r^2+m^2} +\sum^{N}_{i=1}h(\eta_i),
\l{4.10}\ee
where $\sigma (r)$ is the continuous spectrum and $h(\eta)$ is the
soliton energy. Note absence of interaction energy among solitons.

New degrees of freedom $(\x,\eta)(t)$ must obey the equations
(\r{2.24}):

\ba
\dot{\x}_i= \o (\eta_i) -\int dx j(x,t) \f {\pa u_N (x;\x,\eta)}{\pa
\eta_i},
\n \\
\dot{\eta}_i=\int dx j(x,t) \f {\pa u_N (\x, \eta)}{\pa \x_i},~~~ \o
(\eta) \equiv \f {\pa h(\eta)}{\pa \eta}.
\l{4.11}\ea
Hence the source of quantum perturbations are proportional to
the time-local tangent vectors $\pa u_N(x;\x,\eta)/\pa \eta_i$ and
$\pa u_N (x;\x, \eta)/\pa\x_i$ to the soliton configurations. It
suggests the idea to split the `Lagrange' sources: $j(x,t) \rar
(j_{\x}, j_{\eta})$. This leads to new weight functional
$U(u_N;e_{\x}, e_{\eta})$ and new perturbations generating operator
$\hat{K}(e_{\x},e_{\eta}; j_{\x},j_{\eta})$.

In result:
\ba
\R_2(q)= \sum_N e^{-i \hat{K}(e_{\x},e_{\eta};j_{\x},j_{\eta})}
\n\\\times
\int D^N M (\x,\eta)e^{is_0(u_N)}e^{-iU(u_N;e_{\x},e_{\eta})}
|\Ga (q;u_N)|^2
\l{4.13}\ea
where, using vector notations,
\be
\hat{K}(e_{\x},e_{\eta};j_{\x},j_{\eta})
=\f{1}{2}Re\int_{C_+} dt \{\hat{j}_{\x} (t)\cdot \hat{e}_{\x}(t)+
\hat{j}_{\eta}(t)\cdot \hat{e}_{\eta}(t)\}.
\ee
The measure takes the form:
\ba
D^N M (\x,\eta)=\prod^{N}_{i=1}\prod_{t}d\x_i (t) d\eta_i (t)
\n\\\times
\d (\dot{\x}_i- \o (\eta_i) - j_{\x,i}(t))
\d (\dot{\eta}_i - j_{\eta,i}(t))
\l{b}\ea
The effective potential
\be
U(u_N;e_{\x},e_{\eta})=-\f{2m^2}{\la^2}\int dx dt \sin\la u_N~
(\sin \la e -\la e)
\l{4.15}\ee
with
\be
e(x,t)=e_{\x}(t) \cdot \f{\pa u_N (x;\x,\eta)}{\pa \eta (t)}-
e_{\eta}(t) \cdot \f{\pa u_N (x;\x, \eta)}{\pa \x (t)}.
\l{4.15'}\ee

Performing the shifts:
\ba
\x_i (t) \rar \x_i (t) + \int dt' g(t-t') j_{\x,i}(t') \equiv
\x_i (t) +\x'_i (t),
\n \\
\eta_i (t) \rar \eta_i (t) + \int dt' g(t-t') j_{\eta ,i}(t') \equiv
\eta_i (t) +\eta'_i (t),
\l{4.16}\ea
we can get the Green function $g(t-t')$ into the operator
exponent:  \ba
\hat{K}(e_{\x},e_{\eta};j_{\x},j_{\eta}) =\f{1}{2}\int
dt dt' \Th (t-t')\{\hat{\x}'(t')\cdot \hat{e}_{\x}(t)
\n\\
+\hat{\eta}'(t')\cdot \hat{e}_{\eta}(t)\}.
\ea
since the Green function $g(t-t')$ of eqs.(\ref{4.11}) is the step
function:
\be
g(t-t')=\Th (t-t')
\l{4.12}\ee
Its imaginary part is equal to zero for real times and this allows to
shift $C_{\pm}$ to the real-time axis. Note the Lorentz
noncovariantness of our perturbation theory with Green function
(\r{4.12}).

In result:
\be
D^N M(\x,\eta)=\prod^{N}_{i=1}\prod_{t}d\x_i(t) d\eta_i(t)
\d (\dot{\x}_i- \o (\eta+\eta'))\d (\dot{\eta}_i)
\l{c}\ee
with
\be
u_N=u_N (x;\x+\x',\eta+\eta').
\l{4.19}\ee

The equations:
\be
\dot{\x}_i=\o (\eta_i+\eta'_i)
\ee
are trivially integrable. In quantum case $\eta'_i \neq 0$ this
equation describes motion in the nonhomogeneous and anisotropic
manifold. So, the expansion over $(\hat{\x'},~\hat{e}_{\x},~\hat{\eta}',
~\hat{e}_{\eta})$ generates the local in time fluctuations of
$W$ manifold. The weight of this fluctuations is defined by
$U(u_N;e_{\x},e_{\eta})$.

Using the definition:
$$
\int Dx \d (\dot{x})=\int dx(0)=\int dx_0
$$
functional integrals are reduced to the ordinary integrals over initial
data $(\x,\eta)_{0}$. This integrals define zero modes volume. Note
once more that the zero-modes measure was defined without
Faddeev-Popov $anzats$.

The proof of (\r{1}) we would divide on two parts. First of all we would
consider the quasiclassical approximation and then we will show that
this approximation is exact.

This strategy is necessary since it seems to important to show the
role of quantum corrections noting that for all physically
acceptable field theories\footnote{It is natural to assume that the
fields should tend to zero at $\s_\infty$ \C{coll}.} $\R_{nm}=0$ in
the quasiclassical approximation.  We would like to show that\\

{\it S22. The exactness of quasiclassical approximation is the
necessary and sufficient condition to have (\r{1}) in the model
(\r{1.13},~ \r{2.7}).}\\
Note, this statement is not evident from the unitarity condition.

The $N$-soliton solution $u_N$ depends from $2N$ parameters. Half
of them $N$ can be considered as the position of solitons and other
$N$ as the solitons momentum. Generally at $|t|\rightarrow\infty$
the $u_N$ solution decomposed on the single solitons $u_s$ and on
the double soliton bound states $u_b$ \C{takh}:
$$
u_N(x,t)=\sum^{n_1}_{j=1}u_{s,j}(x,t)+\sum^{n_2}_{k=1}u_{b,k}(x,t)+
O(e^{-|t|})
$$
Note that this asymptotic is achieved if $\x_i\rar\infty$ or/end
$\eta_i\rar\infty$. Last one defines the befurcation line of our
model. So, the one soliton $u_s$ and two-soliton bound state $u_b$
would be the main elements of our formalism.  Its $(\x,\eta)$
parametrizations confirmed to eqs.(\r{2.17'},~\r{2.23}) have the
form:
\be
u_s(x;\x,\eta)=-\f{4}{\la}\arctan\{\exp(mx\cosh\b\eta
-\x)\},~~~ \b =\f{\la^2}{8}
\l{so} \ee
and
\ba
u_b(x;\x,\eta)=
\n\\
-\f{4}{\la}\arctan\{\tan\f{\b\eta_2}{2}
\f{mx\sinh \f{\b\eta_1}{2}\cos \f{\b\eta_2}{2}-\x_2}
{mx\cosh \f{\b\eta_1}{2}\sin \f{\b\eta_2}{2}-\x_1}\}.
\l{bo}\ea

The $(\x,\eta)$ parametrization of solitons individual energies $h(\eta)$
takes the form:
$$
h_s(\eta)=\f{m}{\b}\cosh \b\eta,~~~
h_b(\eta)=\f{2m}{\b}\cosh \f{\b\eta_1}{2}\sin\f{\b\eta_2}{2}\geq 0.
$$
The bound-states energy $h_b$ depends from $\eta_2$ amd $\eta_1$.
First one defines inner motion of two bounded solitons and second one
the bound states center of mass motion. Correspondingly we will call
this parameters as the internal and external ones. Note that the
inner motion is periodic, see (\r{bo}).

Performing last integration we find:
\ba
\R_2(q)=
\sum_N \int \prod^{N}_{i=1} \{d\x_0 d\eta_0\}_i
e^{-i \hat{K}}e^{is_0(u_N)}
\n\\\times e^{-iU(u_N;e_{\x},e_{\eta})}
|\Ga (q;u_N)|^2
\l{5.1}\ea
where
\be
u_N=u_N (\eta_0 +\eta',\x_0 + \o (t) +\x').
\l{5.2}\ee
and
\be
\o (t)=\int dt'  \th (t-t') \o (\eta_0 +\eta')(t')
\l{5.3}\ee

In the quasiclassical approximation $\x'=\eta'=0$ we have:
\be
u_N=u_N (x;\eta_0 ,\x_0 + \o (\eta_0)t).
\l{5.7}\ee
Note now that if the surface term
\be
\int dx^\mu\pa_{\mu}(e^{iqx}u_N)=0
\l{5.8}\ee
then
\ba
\int d^2x e^{iqx}(\pa^2 +m^2)u_N (x,t)
\n\\
=-(q^2-m^2)\int d^2x e^{iqx}u_N (x,t) =0
\l{5.9}\ea
since $q^2$  belongs to mass shell by definition. The condition
(\ref{5.8}) is satisfied for all $q_{\mu}\neq0$ since $u_N$ belong
to Schwarz space (the periodic boundary condition for $u(x,t)$ do not
alter this conclusion). Therefore, in the quasiclassical approximation
(\r{1}) is hold.

Expending the operator exponent in (\ref{5.1}) we find that action of
operators $\hat{\x}'$, $\hat{\eta}'$ create terms

\be
\sim\int d^2x e^{iqx} \th (t-t') (\pa^2 +m^2)u_N (x,t) \neq 0
\l{5.12}\ee
This ends our proof.\\

$\bullet$ Now we will show that\\

{\it S23. The quasiclassical approximation is exact in the soliton
sector of sin-Gordon model.}
The structure of the perturbation theory is readily seen in the `normal-
product' form:
\ba
\R_2(q)=\sum_N \int \prod^{N}_{i=1} \{d\x_0 d\eta_0\}_i
\n\\\times
:e^{-iU(u_N;\hat{j}/2i)}e^{is_0(u_N)}|\Ga (q;u_N)|^2:,
\l{6.1}\ea
where
\be
\hat{j}=\hat{j}_{\x}\cdot\f{\pa u_N}{\pa \eta}- \hat{j}_{\eta}\cdot
\f{\pa u_N}{\pa \x}=\O \hat{j}_{X}\f{\pa u_N}{\pa X}
\l{6.2}\ee
and
\be
\hat{j}_{X}=\int dt' \Th(t-t')\hat{X}(t')
\l{6.3}\ee
with $2N$-dimensional vector $X=(\x ,\eta)$. In eq.(\r{6.2}) $\O$ is
the ordinary symplectic matrix.

The colons in (\r{6.1}) mean that the operator $\hat{j}$ should stay to the
left of all functions in the perturbation theory expansion over it. The
structure (\r{6.2}) shows that each order over $\hat{j}_{X_i}$ is proportional
at least to the first order derivative of $u_N$ over conjugate to
$X_i$ variable.

The expansion of (\r{6.1}) over $\hat{j}_{X}$ can be written using $S10$ of
\C{yaph}in the form (omitting the quasiclassical approximation):
\be
\R_2(q)=\sum_N \int \prod^{N}_{i=1} \{d\x_0 d\eta_0\}_i
\{\sum^{2n}_{i=1}\f{\pa}{\pa X_{0i}}P_{X_i}(u_N)\},
\l{6.4}\ee
where $P_{X_i}(u_N)$ is the infinite sum of `time-ordered' polinoms
(see \C{yaph}) over $u_N$ and its derivatives. The explicit form of
$P_{X_i}(u_N)$ is complicated since the interaction potential is
nonpolinomial. But it is enough to know, see (\r{6.2}), that
\be
P_{X_i}(u_N)\sim \O_{ij} \f{\pa u_N}{\pa X_{0j}}.
\l{6.5}\ee

Therefore,
\be
\R_2(q)=0
\l{6.6}\ee
since (i) each term in (\r{6.4}) is the total derivative, (ii) we have
(\r{6.5}) and (iii) $u_N$ belongs to Schwarz space.\\

$\bullet$ We can conclude that the equality (\r{6.6}) is not hold iff
\be
\f{\pa u_N}{\pa X_{0}}\neq 0 ~~at~~X_0 \in \a W_s,
\l{6.7}\ee
where the boundary $\pa W_s$ is the bifurcation line of the invariant
subspace.

In our consideration, in accoradance to our selection rule, the
continuous spectrum contributions are absent since they are realized
on zero measure $\sim \{volume~of~W\}^{-1}$).

\section{Differential measure on $T^*\bar{G}_1\times R^5$}\0 ddd

The general properties of quantum dynamics of the massless scalar
fields in the $$W_O=O(4,2)/O(4)\times O(2)$$ factor space would be
described. We start calculations from the integral:
\be
\R_2(q)=e^{-i\h{K}(j,e)}\int DM(u)|\Ga (q,u)|^2 e^{-iU(u, e)},
\l{g1}\ee
where the Dirac measure
\be
DM(u)=\prod_{x,t} du(x,t) \d (\pa^2 u(x,t) + gu^3(x,t) -j(x,t))
\l{g2}\ee
defines a complete set of contributions and
\be
\Ga (q,u)=\int d^3xdt e^{iqx}\pa_\mu^2 u(x,t).
\l{g3}\ee
for this theory. The expansion of operator exponent $e^{-i\h{K}}$,
\be
\h{K}(j,e)=\f{1}{2}\int d^3x dt\h{j}(x,t)\h{e}(x,t),
\l{g4}\ee
means the expansion in vicinity of zero of auxiliary variables $j$
and $e$.  This allows to start from the unperturbated equation
\be
\pa^2 u + gu^3 =0
\l{g5}\ee
finding the contributions into integral (\r{g1}).

The Green function is singular on the light cone and to avoid this
singularity the $i\e$-prescription should be used.  This imply
continuation of (\r{g1}) into the space of complex fields.
Nevertheless we simplify notations considering
\be
U(u,e)=2g\Re\int_{C_+}d^3x dtu(x)e^3(x)
\l{g6}\ee
to describe the interactions since the mapping into the $W_O$ space
would be considered.\\

$\bullet$ We will take into account following solution of (\r{g5})
(see \C{act} and references cited therein):
\ba
\p(x)=(\f{4}{g\eta_1^2})^{1/2}\{ (1+\f{(x-x_0)^2}
{\eta_1^2})^2
\n\\
+ (2\f{\eta_2l_\mu (x-x_0)^\mu}{\eta_1})^2\}^{-1/2}
=O(1/\sqrt g),
\l{g7}\ea
where $x_0$, $l$ are the 4-vectors. $Note$, this solution of
eq.(\r{g5}) is $O(4)\times O(2)$ invariant \C{act}. It is assumed
in (\r{g7}) that
\be
l_\mu l^\mu=+1/\eta_2^2\geq0,~~\vec{l}^2=1,
\l{g8'}\ee
i.e. $1/\eta_2$ is the time scale. Then $\p$ is the 8 parametric
function.

The $W_O$ space we define by the conditions:
\be
\infty\geq\eta_1^2\geq 0,~|\eta_2|\leq\infty,~|x_{0,\mu}|\leq\infty
\l{g8a}\ee
$Note$, $\vp$ becomes imaginary if $\eta_1^2$ is negative. By this
reason the region $\eta_1^2<0$ would not be considered. Therefore,
\be
\pa W_O:~=\{\eta_1=0,+\infty,~\eta_2=\pm\infty,~x_{0,\mu}=
\pm\infty\}.
\l{g8}\ee
$Note$, The integrals over $\vp(x)$ becomes singular at $\eta_1=0$.
To avoid this unphysical singularities the $i\e$-prescription was
preserved, see (\r{g6}).

$Note$, the $naive$ insertion of (\r{g7}) into (\r{g3}) gives
$\R^{qc}_2(q)=0$ if $q_\mu \neq 0$. It will be shown that the quantum
corrections gets to $\R_2\neq 0$.

The parameter $\eta_1$ defines the scale of $\vp$. Considered solution
(\r{g7}) has following asymptotics:
\be
\p (x)\sim \f{\eta_1}{|x|} ~~at~~\eta_1/|x| \rar 0
\l{g9}\ee
and
\be
\p (x) \sim \f{1}{\eta_1}~~at~~\eta_1/|x| \rar \infty .
\l{g10}\ee
We will see that the asymptotics (\r{g9}) leads to divergences in
the perturbation series. Following to estimations (\r{g9},\r{g10})
$\eta_1=\infty$ defines the `ultraviolet' region over $x$ and the
`infrared' region over $x$ correspond to $\eta_1=0$. Therefore, our
perturbation theory would contain the ultraviolet divergences and be
the infrared stable.

The quantum corrections are defined on the boundary $\pa W_O$, see
S14, and $\R_2(q)\neq0$ if
\be
\{\vp\}\bigcap\pa W_O\neq\emptyset.
\l{g11}\ee
Noting (\r{g7}) and (\r{g8}) one may expect that $\R_2(q)\neq0$ since
\be
\{\vp\}\bigcap\inf\pa_{\eta_1}W_O\neq\emptyset.
\l{g12}\ee
in contrast to Coulomb problem and sin-Gordon model, where the
intersection of `fields' set with boundaries was empty.

{\it Note,} other directions in the $W_O$ space did not give the
contributions since \ba\{\vp\} \bigcap \pa_{\eta_2} W_O=
\emptyset,\n\\\{\vp\}\bigcap \sup \pa_{\eta_1} W_O= \emptyset,~\{\vp\}
\bigcap\pa_{x_{0,\mu}}W_O= \emptyset.\ea The expected nontrivial
result $\R_2(q)\neq0$ we interpret as a consequence of broken scale
symmetry since its appearance is connected with non-emptiness of the
$\pa_{\eta_1}W_O$ boundary.  Certainly, the more careful analysis is
needed for this conclusion.

With definition (\r{g8a}) the classical fields energy:
\ba
h_c (\eta)=\int d^3 x \{ \f{1}{2}(\pa_0
\p)^2 + \f{1}{2}(\vec{\pa} \p)^2 +\f{1}{4}g\p^4\}
\n\\
=\f{1}{\sqrt{|\eta_1^2|}}h(\eta_2)\geq 0
\l{g13}\ea
is the well defined conserved quantity. $Note$, the classical fields
energy $h_c \rar \infty$ at $\eta_1 \rar 0$, but the renormalization
procedure would `hide' this divergences. $Note$, the singularity at
$\eta_1=0$ is the point of bifurcation.

Other properties of $\p$ one can find in \C{act}.\\

$\bullet$ In the phase space the measure (\r{g2}) has the form:
\be
DM(u,p)=\prod_{x,t} du(x,t) dp(x,t) \d(\dot{u}-\f{\d H_j}{\d p})
\d(\dot{p}+\f{\d H_j}{\d u}),
\l{g14}\ee
where the total Hamiltonian
\be
H_j (u,p)=\int d^3 x \{
\f{1}{2} p^2 + \f{1}{2}(\vec{\pa} u)^2 +\f{1}{4}gu^4 -ju\}
\l{g15}\ee
includes the energy of quantum excitations $ju$.

We want to show that\\

{\it S24. The mapping on the $W_O$ manifold gives
\ba
DM(\x, \eta)=d^3x_0 d^3l
\d(\vec{l}^2-1)dt_0 \d(\x_1(0)-\x_2(0))
\n \\
\times\prod_t d^2\x (t)d^2\eta (t) \d^2(\dot{\x}-\f{\pa {h}_j}{\pa \eta})
\d^2(\dot{\eta}+\f{\pa {h}_j}{\pa \x})
\l{g16}\ea
if $u=u_c$, where $(l_0=\pm\sqrt{1+1/\eta_2^2})$
\ba
u_c(\vec{x};\x, \eta)=\f{2\eta_1 {\o}_1^2{\o}_2}{\sqrt{g}}
\{{\o}_2^2 (\eta_1^2 {\o}_1^2 +\x_1 - {\o}_1^2(\vec{x}-\vec{x}_0)^2)^2
\n \\
+4{\o}_1^4\eta_1^2(l_0\x_2
-\eta_2{\o}_2\vec{l}\cdot(\vec{x}-\vec{x}_0))^2\}^{-1/2},
\l{g17}\ea}
Here $h_j$ is the transformed total Hamiltonian:
\be
h_j(\x, \eta)=H_j(u_c,p_c)=h_c(\eta) -\int d^3x
ju_c,~~p_c=\dot{u}_c.
\l{g18}\ee
and the `velocities'
\be
{\o}_i (\eta_1, \eta_2)=\f{\pa}{\pa \eta_i}{h}_c(\eta_1, \eta_2).
\l{g19}\ee
was introduced.

The function $u_c$, defined in (\r{g17}), is the solution of incident
eq.(\r{g5}) if ($\x,~\eta$) obey the equations
\be
\dot{\x}_i=\f{\pa {h}_j}{\pa
\eta_i},~~\dot{\eta}_i=-\f{\pa {h}_j}{\pa \x_i}, ~~i=1,2,
\l{g20}\ee
and the boundary condition
\be
\x_1(0)=\x_2(0)
\l{g21}\ee
is applied. One can check this easily inserting into (\r{g17}) the
solutions of eq.(\r{g20}). The same is seen solving the equation
\be
\{u_c, {h}_j\}=\f{\d H_j(u_c,p_c)}{\d p_c},~~p_c=\dot{u}_c,
\l{g21'}\ee
where $\{,\}$ is the Poisson bracket in the $(\x,~\eta)$ phase space.

The proof of $S24$ repeats calculation of measures for Coulomb
problem and sin-Gordon model. But it must be shown to find (\r{g17})
how the constraints can be included into formalism.

Let us consider for this purpose
\ba
\D_F(u,p)=\int\prod_t d\x(t)d\eta(t)
\n\\\times
\prod_{\vec{x},t}\d(u(\vec{x},t)-u_c(\vec{x};\x,\eta))
\d(p(\vec{x},t)-p_c(\vec{x};\x,\eta))
\n\\\times
F(\x,\eta)
\l{g22}\ea
with some known functions $(u_c,~p_c)$ of two $n$-dimensional
vectors $(\x,~\eta).$ The functional $F(\x,\eta)$ was introduced to
take into account the constraints. We will specify it below.

It is evident that $\D_F \sim \prod_x \d$-functions, i.e. is the
distribution. So, it must be defined on the `good' (Schwartz)
support. This means that the continuum of equations
$$
u(\vec{x},t)=u_c(\vec{x};\x,\eta),~~p(\vec{x},t)=p_c(\vec{x};\x,\eta)
$$
should have, independent from $\vec{x}$, $2n$ nontrivial solutions
for $(\x,\eta)$. Then, expanding arguments of $\d$-functions near this
solutions $(\x_c,\eta_c)$ we find:
\ba
\D_F (u,p)=F(\x,\eta)\int \prod_t d\tilde{\x} d\tilde{\eta}
\n\\\times
\prod_{\vec{x},t}
\d(\f{\pa u_c}{\pa \x}\cdot \tilde{\eta} +\f{\pa u_c}{\pa \x}\cdot
\tilde{\eta}) \d(\f{\pa p_c}{\pa \x}\cdot \tilde{\x} +\f{\pa p_c}{\pa
\eta}\cdot \tilde{\eta}).
\l{g23}\ea
Deriving this equality we assume that the functions $(u_c, p_c)$ obey
the condition
\be
\det(\pa u_c, \pa p_c)\neq 0.
\l{g24}\ee
This means that $\D_F \sim \prod_{\vec{x}}\d (0)$. To regularize this
quantity one can divide the $\vec x$ space onto $n$ cells \C{takh}.
One can consider also the $\d$-functions of (\r{g23}) as the limit of
Gauss distribution functions, $$ \d (...)=\lim_{\e =0}\d_{\e}(...) $$
and put $\e =0$ at the very end of calculations. $Note$, $(\x,\eta)$ in
(\r{g23}) are arbitrary functions of $t$.

Inserting
\ba
1=\f{1}{\D_F (u,p)}
\int\prod_t d\x(t)d\eta(t)\n\\\times
\prod_{\vec{x},t}\d(u-u_c)\d(p-p_c)F(\x,\eta),
\ea
where $\D_F (u,p)$ was defined in (\r{g23}), and integrating over
$(u,p)$ we find the measure:
\be
DM(\x,\eta)=\prod_t d\x(t)d\eta(t) \f{F(\x,\eta)}{F(\x_c,\eta_c)} \d
(\dot{\x}-\f{\pa {h}_j}{\pa\eta}) \d(\dot{\eta}-\f{\pa {h}_j}{\pa\x})
\l{g25}\ee
if $u_c,~p_c$ are defined by equations:
$$
\{u_c, {h}_j\}=\f{\d H_j}{\d p_c},~~\{p_c, {h}_j\}=-\f{\d H_j}{\d u_c}
$$
and $h_j$ is the total transformed Hamiltonian.

It is important to note that the constraint term
\be
(F(\x,\eta)/F(\x_c,\eta_c))\equiv (F/F_c)
\l{g26}\ee
was factorized in (\r{g25}).\\

$\bullet$ The last step based on the identity \C{yaph}:
$$
\prod_t \d (X-j_x)=e^{-\f{i}{2}\int dt \h{j}'_x
\h{e}_x}e^{2i \int dt e_x j_x}
\d (X-j'_x).
$$
Therefore,
\ba
DM=e^{-\f{i}{2}\int dt (\h{j}_\x\cdot\h{e}_\x
+\h{j}_\eta\cdot\h{e}_\eta)} e^{2i \int dt (\d u_c\wedge\d p_c)} \n
\\ \times\prod_t d^n\x d^n\eta \d^{(n)}(\dot{\x}-\f{\pa{h_c}}{\pa
\eta}-j_\x) \d^{(n)}(\dot{\eta}+\f{\pa {h_c}}{\pa\x}-j_\eta),
\l{g27}\ea
where
\be
\d u_c\wedge\d p_c= e_\x (t)\cdot \f{\pa u_c}{\pa
\eta(t)}- e_\eta (t)\cdot \f{\pa u_c}{\pa \x(t)}
\l{g28}\ee
reflects the symplectic structure of the transformed phase space.

We can calculate action of the operator $\exp\{-i\h{K}(j,e)\}$
and in result, extracting new perturbations generating operator
$\exp\{-i\h{K}_t(j,e)\}$, where
\be
\h{K}_t (j,e)=\f{1}{2}\int dt (\h{j}_\x\cdot\h{e}_\x
+\h{j}_\eta\cdot\h{e}_\eta),
\l{g29}\ee
we find the measure of transformed theory
\ba
DM(X, Y)=\prod_t d^n \x(t) d^n \eta(t) \f{F}{F_c}
\n\\\times
\d(\dot{\x}-\f{\pa {h_c}}{\pa \eta} -j_\x)
\d(\dot{\eta}-\f{\pa {h_c}}{\pa\x} -j_\eta).
\l{g30}\ea
At the same time we should change in (\r{g6})
\be
e(x,t)\rar e_c(x,t)=\d u_c\wedge \d p_c.
\l{g31}\ee
The resulting theory describes perturbations in the $(\x,~\eta)$
phase space.

To adopt the general definitions (\r{g29}-\r{g31}) to our concrete
problem note that our solution extracts two generators, $T_0$ and
$K_0$. Therefore, for invariant subspace definition we will choose,
in accordance with (\r{g16}), $\x_i,~~\eta_i,~~i=1,2.$ Other
coordinates can be chosen arbitrarily. For instance,
$$\x_{2+i}=x_{0i},~\eta_{2+1}=0,~\x_{5+i}=l_i,~\eta_{5+i}=0,~i=1,2,3$$
and other $(\x,~\eta)=0$. This means that $u_c =u_c
(\vec{x}-\vec{x}_0;\x, \eta,\vec{l})$ is $\eta_i,~i=3,4,...,$
independent.  Then
\be
\h{K}_t=\int dt (\h{j}_\x\cdot \h{e}_\x
+\h{j}_\eta\cdot \h{e}_\eta)
\l{g32}\ee
since there is not canonically conjugate pare for $\x_i,~i=3,4,..$.
Taking into account the definition:
$$
\prod_t dX(t) \d (\dot{X})=dX(0)\equiv dX_0.
$$
corresponding measure take the form:
\be
DM= d^3x_0d^3l\prod_t d^2\x d^2\eta
\f{F}{F_c}\d^2 (\dot{\x}-{\o}-j_\x)\d^2 (\dot{\eta}-j_\eta)
\l{g33}\ee
with
\be
e_c(x,t)=e_\x(t)\cdot\f{\pa u_c(x,t)}{\pa
\eta(t)}-e_\eta\cdot\f{\pa u_c(x,t)}{\pa \x(t)}.
\l{g34}\ee
So, the invariant subspace $T^*\bar{G}$ only, with local coordinates
$(\x,~\eta)$, is influenced by quantum perturbations.

The measure (\r{g33}) contains 10 degrees of freedom. But only 8
among them are independent. So, we should shrink the space with
measure (\r{g33}) on two units. For this purpose we would use extra
factor $F(\x,\eta)$ in (\r{g22}) choosing, for instance,
$$
F(\x,\eta)=\d(\sum_i\x_{5+i}^2-1)\d(\x_1(0)-\x_2(0))
$$
Then (see Appendix E)
\be
\f{F}{F_c}=\d(\vec{l}^2-1)\d(\x_1(0)-\x_2(0))
\l{g35}\ee
We would consider $|\x_i|\leq\infty$. Therefore,
\be
\{u_c\}\bigcap\pa_\x W_O=\emptyset.
\l{g35'}\ee

This completes splitting of $W_O$ onto $T^*\bar{G}_1\times M_5$
invariant subspace. $Note$, the $x$ dependence is practically
disappeared and the reduced problem looks like quantum
mechanical one. This property of our approach reflects the
Lorentz-noncovariantness of developed perturbation theory.

One can use the perturbation theory in terms of the initial
`Lagrange' source $j$ and conjugate to it auxiliary field $e$. In
this case the formalism is manifestly Lorentz-covariant. But this
formulation is `rough', it unables to take into account the topology
of the $W$ space.\\

$\bullet$ Further main results are the consequence of the statement
that each term of the perturbation theory can be written as the total
derivative over $(\x,~\eta)$.

We found that
\be
\R(q)=e^{-\f{i}{2}\h{K}_t(j,e)}\int DM |\Ga (q,u_c)|^2
e^{-i V(u_c, e_c)},
\l{g36}\ee
where $\h{K}_t$ was defined in
(\r{g32}), $DM$ in (\r{g33}), (\r{g35}), $V$ in (\r{g6}) with $e_t$
defined in (\r{g34}) and the function $u_c$ is given in (\r{g17}).

Last integrations over $(\x,~\eta)$ in (\r{g36}) are trivial. To
perform them we can use the shift:
\be
\x \rar \x_c=\x_0 + {\o} + \x,~~\eta \rar \eta_c=\eta_0 +\eta,
\l{g37}\ee
where $(\x,~\eta)$ in the r.h.s. are solutions of equations:
\be
\dot{\x}=j_\x,~~\dot{\eta}=j_\eta.
\l{g38}\ee
The Green function of this equations $g(t-t')$ is ordinary step
function $\th (t-t')$, see S.

The shift (\r{g37}) gives the nonlocal operators $\hat{j}_\ga$:
\be
\h{j}_\ga(t)=\int dt' g(t-t') \h{\ga}(t'),
\l{g39}\ee
where $\h\ga=(\h\x,~\h\eta)$ is the four-vector and
\be
\h{K}_t=\f{1}{2}\int dt dt' \th(t-t') \h{e}(t)\cdot \h{\ga} (t')
\l{g40}\ee
with four-vector $\h e=(\h e_\x,~\h e_\eta)$.

The remaining integrations over $(\x_0,~\eta_0)$ are ordinary
integrals with measure
\be
dM= d^3x_0d^3ld^2\x_0 d^2\eta_0 \d (\x_{c1}(0)
-\x_{c2}(0))\d(\vec{l}^2-1)
\l{g41}\ee

Introducing $\exp\{-i\h{K}\}$ into the integral:
\be
\R(q)=\int dM e^{\h{K}} |\Ga (q;u_c)|^2 e^{-U(u_c;e_c)}
\l{g44}\ee
(the trivial rotation $e\rar -ie$, $\h{e}\rar i\h{e}$ was performed) we can
calculate action of the operator $\exp\{-\h{K}\}$ at $\ga =0$.

The operator $\h{K}$ is linear over $\h{e}$. Therefore,
\be
\R (q)=\int dM :e^{\h{U}(u_c, j)/4} |\Ga(q;u_c)|^2:,
\l{g45}\ee
where
\be
\h{U}(u_c, j)=2g\Re\int_{C_+}dxdt (\d\h{u_c}\wedge \d p_c)^3u_c
\l{g46}\ee
where
\be
\d\h{u_c}\wedge \d p_c=\h j_\ga\wedge \d p_c
\l{g47}\ee
and $\h j_\ga$ was defined in (\r{g39}). The colons mean normal
product when the operators stay to the left of functions.

Calculating the integral over $x$ in (\r{g46}) we reduce our
field-theoretical problem to the quantum-mechanical one with
complicated non-polinomial potential of interactions. We plan to
consider this perturbation series for more realistic Yang-Mills
theory.

We conclude this section noting that, using $S$ and $S$, that
$\R(q)\neq0$ since
$$\{u_c\}\bigcap\inf\pa_{\eta_1} W_O\neq\emptyset.$$

\section{Instead of conclusion}\0

Described approach is based on three `whales'. They are (i) the
definition of observables in quantum theories as the modulo square of
amplitudes, (ii) the description of quantum processes as the
transformation induced by unitary operator $\exp\{iS(x)\}$, where
$S(x)$ is the classical action and (iii) the unitarity condition as
the principle which determines connection between quantum dynamics
and classical measurement. Less principal assumption, usually taken
`by treaty', that the quantum perturbations are switched on
adiabatically was used also.\\

$\bullet$ To use all above fundamental principles the formalism in
terms of observables only was considered. We tryed to show that such
approach is sufficiently general being able to describe a wide
spectrum of experiments (without claim on general philosophy). Our
approach should be considered as the useful technical trick (probably
not unique) helping to calculate the observables if the nontrivial
topologies should be taken into account and the corresponding
physical vacuum is so complicated that its definition is hopeless
task.

Following points of the formalism should be picked out. Firstly,
wishing to count quantum perturbations of fields we generalize on
quantum case the postulate of classical mechanics that the initial
conditions $(\x_0,\eta_0)$ determines the classical trajectory
$\vp$ completely, $\vp=\vp(\vec{x},t;\x_0,\eta_0)$. In considered
examples the set of this parameters form the manifold $W_G$. Then we
introduce the dynamics in the $W_G$ space noting that the quantum
motion in the $W_G$ space should describe all excitations of $\vp$
since the complet set of field states  $\{\vp\}\in W_G$. The
unitarity of used mapping guaranteed by $\d$-likeness of measure.
This, new for quantization schemes trick, significantly simplify
perturbation theory.

Secondly, we discovered that the motion in definite directions of the
$W_G$ space is exactly classical. This is the new phenomena in
quantum theories. We found that $W_G$ may decoupled on direct product
of $T^*G$ subspace and $R^n$. The quantum dynamics is realized in
the symplectic $T^*G$ subspace of the $W_G$ space and motion in
$R^n$is exactly classical. Actually we found the generalization of
the canonical quantization scheme.

Thirdly, the analyses of the perturbation theory allows to show
that the quantum corrections are accumulated on the boundary $\pa
W_G=\pa T^*G$. Then, if $\{\pa\vp\}\bigcap\pa T^*G=\emptyset$ the
quasiclassical approximation is exact. We would assume this
phenomena as a basis of the confinement since in this case the
particle creations generating functional is trivial. This formulation
seems selfconsistent because of $\d$-likeness of measure and if
the above selection rule taken into account.

\vspace{0.2in}
{\Large \bf Acknowledgement}

We would like to thank V.G.Kadyshevski to the helpful discussions.
The paper was compiled as the result of kind interest of M.Saveliev.
One of us (JM) was partly granted by Georgian Academy of Sciences.

\renewcommand{\theequation}{A.\arabic{equation}}

\appendix\section{Appendix}\0

In the CM frame we have:
\ba
n_{++}(q_0)=n_{--}(q_0)=
\frac{\sum_{n=0}^{\infty}ne^{-\frac{\beta_1+\beta_2}{2}|q_0|n}}
{\sum_{n=0}^{\infty}e^{-\frac{\beta_1+\beta_2}{2}|q_0|n}}=
\n \\
=\frac{1}{e^{\frac{\beta_1 +\beta_2}{2}|q_0|}-1}=
\tilde {n}(|q_0|\frac{\beta_1 +\beta_2}{2}).
\l{47'}\ea
Computing $n_{ij}$ for $i\neq j$ we must take into account that we
have one more particle:
\ba
n_{+-}(q_0)=
 \theta (q_0)
\frac{\sum_{n=1}^{\infty}ne^{-\frac{\beta_1+\beta_1}{2}q_0 n}}
{\sum_{n=1}^{\infty}e^{-\frac{\beta_1+\beta_1}{2}q_0 n}}
\n\\
+ \Theta (-q_0)
\frac{\sum_{n=0}^{\infty}ne^{\frac{\beta_1+\beta_1}{2}q_0 n}}
{\sum_{n=0}^{\infty}e^{\frac{\beta_1+\beta_1}{2}q_0 n}}=
\n \\
= \Theta (q_0)(1+\tilde {n}(q_0 \beta_1))+
 \Theta (-q_0)\tilde {n}(-q_0 \beta_1)
\l{48'}\ea
and
\be
n_{-+}(q_0)=
 \Theta (q_0)\tilde {n}(q_0 \beta_2)+
 \Theta (-q_0)(1+ \tilde {n}(-q_0 \beta_2)).
\l{49'}\ee
Using (\r{47'}-\r{49'}) we find the Green functions ($z=1$):
\be
G_{i,j}(x-x',\b)=\int \frac{d^4 q}{(2\pi)^4} e^{iq(x-x')}
\tilde{G}_{ij} (q, \b)
\l{50'}
\ee
where
\ba
i\tilde{G}_ij (q, \b)=
\left( \matrix{
\frac{i}{q^2 -m^2 +i\epsilon} & 0 \cr
0 & -\frac{i}{q^2 -m^2 -i\epsilon} \cr
}\right)
+\n \\ \n \\+
2\pi \delta (q^2 -m^2 )
\left( \matrix{
\tilde{n}(\frac{\beta_1 +\beta_2}{2}|q_0 |) &
\tilde{n}(\beta_2 |q_0 |)a_+ (\beta_2) \cr
\tilde{n}(\beta_1 |q_0 |)a_- (\beta_1) &
\tilde{n}(\frac{\beta_1 +\beta_2}{2}|q_0 |) \cr
}\right)
\l{51'}
\ea
and
\be
a_{\pm}(\beta)=-e^{\frac{\beta}{2}(|q_0|\pm q_0)}.
\l{52'}
\ee

\renewcommand{\theequation}{B.\arabic{equation}}
\section{Appendix}\0

To show the splitting mechanism let us consider the action of the
perturbation - generating operators:
\ba e^{-i\f{1}{2}\Re\int_{C_+}
dt \hat{j}(t)\hat{e}(t) }e^{-iU_T (x_c,e)}
\n\\\times
\prod_{t} \d (\dot{h}
-j\frac{\pa x_c}{\pa \th}) \d (\dot{\th} -1 +j\frac{\pa x_c}{\pa h}=
\n \\ =\int De_h De_{\th}e^{2i\Re\int_{C_+}dt (e_h
\dot{h}+e_{\th}(\dot{\th}-1))} e^{-iU_T (x_c,e_c )},
\l{22'''}\ea
where
\be e_c =e_h \frac{\pa x_c}{\pa \th} -e_{\th} \frac{\pa x_c}{\pa h}
\equiv (e_h \h{\th}-e_\th\h{h})x_c.  \l{23'''}\ee
The integrals  over
$(e_h ,e_{\th})$ will be calculated perturbatively:  \ba e^{-iU_T
(x_c,e_c )}=\sum^{\infty}_{n_h ,n_{\th} =0}\frac{1}{n_h !n_{\th} !}
\n\\\times
\int \prod^{n_h }_{k=1}(dt_k e_h (t_k))\prod^{n_{\th} }_{k=1}(dt'_k
e_{\th} (t'_k)) \n \\\times P_{n_h ,n_{\th}} (x_c ,t_1
,...,t_{n_h},t'_1,...,t_{n_{\th}}), \l{24''} \ea where \ba P_{n_h
,n_{\th}} (x_c ,t_1 ,...,t_{n_h},t'_1,...,t_{n_{\th}})=
\n\\\times\prod^{n_h
}_{k=1} \hat{e}'_h (t_k) \prod^{n_{\th} }_{k=1} \hat{e}'_{\th}
(t'_k)e^{-iU_T (x_c,e'_c )} \l{25'''} \ea with $e'_c \equiv e_c (e'_h
,e'_{\th} )$ and the derivatives in this equality are calculated at
$e'_h =0$, $e'_{\th}=0$. At the same time, \ba \prod^{n_h }_{k=1} e_h
(t_k)\prod^{n_{\th} }_{k=1} e_{\th} (t'_k)= \prod^{n_h }_{k=1}
(i\hat{j}_h (t_k))\prod^{n_{\th} }_{k=1} (i\hat{j}_{\th} (t'_k)) \n
\\ \times e^{-2i\Re\int_{C_+} dt (j_h (t)e_h
(t)+j_{\th}(t)e_{\th}(t))}.  \l{26'''}\ea The limit $(j_h ,
j_{\th})=0$ is assumed. Inserting (\ref{25'''}), (\ref{26'''}) into
(\ref{22'''}) we find new representation for $R(E)$:  \ba \R(E)=2\pi
\int^{\infty}_{0}
dTe^{\frac{1}{2i}(\hat{\omega}\hat{\tau}+\Re\int_{C_+} dt (\hat{j}_h
(t)\hat{e}_h (t)+ \hat{j}_{\th} (t)\hat{e}_{\th} (t)))}
\n\\\times\int Dh D\th
e^{-i\tilde{H}(x_c ;\tau )-iU_T (x_c ,e_c )} \n\\ \times \delta (E+
\omega -h(T))\prod_{t} \delta (\dot{h} -j_h )\delta (\dot{\th} -1 -
j_{\th}) \l{27''}\ea in which the `energy' and the `time' quantum
degrees of freedom are splitting.

\renewcommand{\theequation}{C.\arabic{equation}}
\section{Appendix}\0

By definition $U_T$ is the odd over $\h{e}_c$ local functional:  \be
V_T (x_c ,\h{e}_c )=2\int^T _0 \sum^{\infty}_{n=1} (\h{e}_c
(t)/2i)^{2n+1} v_n (x_c), \l{a4}\ee where $v_n (x_c)$ is some
$function$ of $x_c$. Inserting (\r{a2}) we find:  \be :e^{-iV_T (x_c
,\h{e}_c )}:=\prod^{\infty}_{n=1}\prod^{2n+1}_{k=0}
:e^{-iV_{k,n}(\h{j},x_c)}:, \l{a5}\ee where \be
V_{k,n}(\h{j},x_c)=\int^{T}_{0}dt (\h{j}_{\p}(t))^{2n-k+1}
(\h{j}_{I}(t))^{k}b_{k,n}(x_c).  \l{a6}\ee Explicit form of the
function $b_{k,n}(x_c)$ is not important.

Using definition (\r{a3}) it easy to find:  $$ \h{j}(t_1)b_{k,n}(x_c
(t_2))=\Th (t_1 -t_2) \pa b_{k,n}(x_c)/\pa X_0 $$ since $x_c =x_c
(X(t)+X_0)$, see (\r{3.33'}), or \be
\h{j}_{X,1}b_2=\Th_{12}\pa_{X_0}b_2 \l{a7}\ee since indices $(k,n)$
are not important.

Let as start consideration from the first term with $k=0$. Then
expanding $\h{V}_{0,n}$ we describe the angular quantum fluctuations
only. Noting that $\pa_{X_0}$ and $\h{j}$ commute we can consider
lowest orders over $\h{j}$. The typical term of this expansion is
(omitting index $\p$) \be \h{j}_1 \h{j}_2 \cdots \h{j}_m b_1 b_2
\cdots b_m.  \l{a8}\ee It is enough to show that this quantity is the
total derivative over $\p_0$.  The number $m$ counts an order of
perturbation, i.e. in $m$-th order we have $(\h{V}_{0,n})^m$.

$m=1$. In this approximation we have, see (\r{a7}), \be \h{j}_1 b_1 =
\Th_{11}\pa_0 b_1 =\pa_0 b_1 \neq 0.  \l{a9}\ee Here the definition
(\r{3.24a}) was used.

$m=2$. This order is less trivial:  \be \h{j}_1 \h{j}_2 b_1 b_2
=\Th_{21} b^2_1 b_2 + b^1_1 b^1_2 +\Th_{12}b_1 b^2_2, \l{a10}\ee
where \be b^n_i \equiv \pa^n b_i.  \l{a11}\ee Deriving (\r{a10}) the
first equality in (\r{3.24c}) was used. At first glance (\r{a10}) is
not the total derivative. But inserting $$ 1=\Th_{12}+\Th_{21},$$
(see the second equality in (\r{3.24c})) we can symmetrize it:  \ba
\h{j}_1 \h{j}_2 b_1 b_2 =\Th_{21}( b^2_1 b_2 + b^1_1 b^1_2)+
\Th_{12}(b_1 b^2_2+b^1_1 b^1_2) \n \\ =\pa_0 (\Th_{21} b^1_1 b_2
+\Th_{12}b_1 b^1_2) \l{a12}\ea since the explicit form of function
$b$ is not important. So, the second order term can be reduced to the
total derivative also. Note, that the contribution (\r{a12}) contains
the sum of all permutations. This shows the `time reversibility' of
the constructed perturbation theory.

Let us consider now expansion over $\h{V}_{k,m}$, $k\neq 0$. The
typical term in this case is \be \h{j}^1_1 \h{j}^1_2 \cdots \h{j}^1_l
\h{j}^2_{l+1} \h{j}^2_{l+2} \cdots \h{j}^2_m b_1 b_2 \cdots b_m ,~~~
0<l<m, \l{a16}\ee where, for instance, $$ \h{j}^1_k \equiv \h{j}_I
(t_k),~~~\h{j}^2_k \equiv \h{j}_{\p} (t_k)$$ and \be \h{j}^i_1 b_2
=\Th_{12}\pa^i_0 b_2 \l{a17}\ee instead of (\r{a7}).

$m=2$, $l=1$. We have in this case:  \ba \h{j}^1_1 \h{j}^2_2 b_1 b_2
= \Th_{21}(b_2 \pa^1_0 \pa^2_0 b_1 + (\pa^2_0 b_2)(\pa^1_0 \pa^2_0
b_1)) \n\\+\Th_{12}(b_1 \pa^1_0 \pa^2_0 b_2 + (\pa^2_0 b_2)(\pa^1_0
\pa^2_0 b_1)) \n\\=\pa^1_0 (\Th_{21} b_2 \pa^2_0 b_1 + \Th_{12}b_1
\pa^2_0 b_2)\n\\ +\pa^2_0 (\Th_{21} b_2 \pa^1_0 b_1 + \Th_{12}b_1
\pa^1_0 b_2).  \l{a18}\ea Therefore, we have the total-derivative
structure yet.

This important property of new perturbation theory is conserved in
arbitrary order over $m$ and $l$ since the time-ordered structure
does not depend from upper index of $\h{j}$, see (\r{a17}).

\renewcommand{\theequation}{D.\arabic{equation}}
\section{Appendix}\0

The resulting measure looks as follows:
\ba
DM(\x, \eta)=\f{1}{\D_c}\d (E-H_0)\prod_t d^2\x d^2\eta
\d (\dot{r_c}-\f{\pa H_j}{\pa p_c})\times
\n \\ \times
\d (\dot{p_c}+\f{\pa H_j}{\pa r_c})
\d (\dot{\vp_c}-\f{\pa H_j}{\pa l_c})
\d (\dot{l_c}+\f{\pa H_j}{\pa \vp_c}),
\l{29}\ea
Note that the parametrization $(r_c,p_c,\vp_c,l_c)(\x,\eta)$ was not
specified.

A simple algebra gives:
\ba
DM(\x, \eta)=\f{\d(E-H_0)}{\D_c}\prod_td^2\x d^2\eta
\int \prod_t d^2\bar{\x}d^2\bar{\eta}
\n \\ \times
\d^2(\bar{\x}-(\dot{\x}-\f{\pa h_j}{\pa\eta}))
\d^2(\bar{\eta}-(\dot{\eta}+\f{\pa h_j}{\pa \x}))
\n \\ \times
\d(\f{\pa r_c}{\pa\x}\cdot\bar{\x}+\f{\pa r_c}{\pa\eta}
\cdot\bar{\eta} +
\{r_c,h_j\}-\f{\pa H_j}{\pa p_c})
\n\\\times
\d(\f{\pa p_c}{\pa\x}\cdot\bar{\x}+\f{\pa p_c}{\pa\eta}
\cdot\bar{\eta} +
\{p_c,h_j\}+\f{\pa H_j}{\pa r_c})
\n \\ \times
\d(\f{\pa \vp_c}{\pa\x}\cdot\bar{\x}+\f{\pa \vp_c}{\pa\eta}
\cdot\bar{\eta}+
\{\vp_c,h_j\}-\f{\pa H_j}{\pa l_c})
\n \\\times
\d(\f{\pa l_c}{\pa\x}\cdot\bar{\x}+\f{\pa l_c}{\pa\eta}
\cdot\bar{\eta}+
\{l_c,h_j\}+\f{\pa H_j}{\pa \vp_c}).
\l{210}\ea
The Poisson notation:
$$
\{X,h_j\}=\f{\pa X}{\pa \x}\f{\pa h_j}{\pa \eta}-
\f{\pa X}{\pa \eta}\f{\pa h_j}{\pa \x}
$$
was introduced in (\r{210}).

We will define the `auxiliary' quantity $h_j$ by following
equalities:
\ba
\{r_c,h_j\}-\f{\pa H_j}{\pa p_c}=0,~
\{p_c,h_j\}+\f{\pa H_j}{\pa r_c}=0,
\n \\
\{\vp_c,h_j\}-\f{\pa H_j}{\pa l_c}=0,~
\{l_c,h_j\}+\f{\pa H_j}{\pa \vp_c}=0.
\l{211}\ea
Then the functional determinant $\D_c$ is canceled and
\be
DM(\x, \eta)=\d(E-H_0)\prod_td^2\x d^2\eta
\d^2(\dot{\x}-\f{\pa h_j}{\pa\eta})
\d^2(\dot{\eta}+\f{\pa h_j}{\pa \x}),
\l{212}\ee

\renewcommand{\theequation}{E.\arabic{equation}}
\section{Appendix}\0

Let us consider
\ba
\D_F(u,p)=\int\prod_t d\x(t)d\eta(t) \prod_{\vec{x},t}
\d(u(\vec{x},t)-u_c(\vec{x};\x,\eta))
\n\\
\times\d(p(\vec{x},t)-p_c(\vec{x};\x,\eta))F(\x,\eta)
\l{e22}\ea
more carefully. Let $(u_c,p_c)(\vec{x};\x,\eta)$ are the known
functions (\r{g17}) of $(\x,\eta)(t)$ and let us assume that $(\x,
\eta)(t)$ are solutions of equations:
\be
\dot{\x}=\pa
h_j/\pa\eta,~\dot{\eta}=-\pa h_j/\pa\x,
\l{e20}\ee
with boundary condition:
\be
\x_1(0)=x_2(0),
\l{e1}\ee
where
\be
h_j(\x, \eta)=H_j(u_c,p_c)=h_c(\eta) -\int d^3x
ju_c,~~p_c=\dot{u}_c.
\l{e18}\ee
The boundary condition (\r{e1}) is necessary to have the 8 parametric
solution of incident equation.

It is assumed in (\r{e22}) that one may find such functions
$(u,p)(x,t)$ that $\D_F(u,p)\neq0$. But we always can invert the
problem saying that $(u,p)(x,t)$ are chosen in such a way that the
condition (\r{e1}) is hold. So, assuming that $(\x,\eta)_c(t)$ obey
the (\r{e1}) condition, we can put
\be
F(\x_c,\eta_c)\equiv F_c=1
\l{e2}\ee

To perform the mapping one should insert
\be
1=\f{1}{\D_{F_c}}
\int\prod_t d\x(t)d\eta(t) \prod_{\vec{x},t}\d(u-u_c)\d(p-p_c)F(\x,\eta),
\l{e3}\ee
where $\D_{F_c}$ was defined in (\r{e22}) with condition (\r{e2}) and
choosing
\be
F(\x,eta)=\d(x_1(0)-x_2(0))\d(\vec{l}^2-1)
\l{e4}\ee
we come to (\r{g35}). Note that the ratio (\r{e3}) equal to one since
$(u,p)_c(x;\x,\eta)$ are obey the equations (\r{g21}). The variables
$(\x,\eta)(t)$ should obey the eqs.(\r{e20}) and the boundary
condition is fixed by (\r{e4}).

\end{document}